\documentclass[aps,twocolumn,preprintnumbers,amsmath,amssymb,nofootinbib,superscriptaddress,notitlepage,10pt]{revtex4-1}

\usepackage{txfonts}
\usepackage{epsfig}
\usepackage{slashed}
\usepackage{color}

\usepackage[utf8]{inputenc}

\begin{document}

\title{Heavy-quark spin and flavour symmetry partners of
  the $X(3872)$ revisited: \\
  what can we learn from the one boson exchange model?}

\author{Ming-Zhu Liu}
\affiliation{School of Physics and Nuclear Energy Engineering, \\
  Beihang University, Beijing 100191, China}

\author{Tian-Wei Wu}
\affiliation{School of Physics and Nuclear Energy Engineering, \\
  Beihang University, Beijing 100191, China}

\author{Manuel Pavon Valderrama}\email{mpavon@buaa.edu.cn}
\affiliation{School of Physics and Nuclear Energy Engineering, \\
Beihang University, Beijing 100191, China} 
\affiliation{
International Research Center for Nuclei and Particles in the Cosmos and \\
Beijing Key Laboratory of Advanced Nuclear Materials and Physics, \\
Beihang University, Beijing 100191, China} 

\author{Ju-Jun Xie}
\affiliation{Institute of Modern Physics, Chinese Academy of
  Sciences, Lanzhou 730000, China}

\author{Li-Sheng Geng}\email{lisheng.geng@buaa.edu.cn}
\affiliation{School of Physics and Nuclear Energy Engineering, \\
Beihang University, Beijing 100191, China} 
\affiliation{
International Research Center for Nuclei and Particles in the Cosmos and \\
Beijing Key Laboratory of Advanced Nuclear Materials and Physics, \\
Beihang University, Beijing 100191, China}

\date{\today}


\begin{abstract} 
  \rule{0ex}{3ex}
  Heavy-quark symmetry as applied to heavy hadron systems implies
  that their interactions are independent of their heavy-quark
  spin (heavy-quark spin symmetry) and heavy flavour
  contents (heavy flavour symmetry).
  In the molecular hypothesis the $X(3872)$ resonance is
  a $1^{++}$ $D^*\bar{D}$ bound state.
  If this is the case, the application of heavy-quark symmetry
  to a molecular $X(3872)$ suggests the existence of a series of
  partner states, the most obvious of which is a possible $2^{++}$ $D^*\bar{D}^*$
  bound state for which the two-body potential is identical to
  that of the $1^{++}$ $D^*\bar{D}$ system, 
  the reason being that these two heavy hadron-antihadron states
  have identical light-spin content.
  As already discussed in the literature this leads to the prediction of
  a partner state at $4012\,{\rm MeV}$, at least in the absence of other
  dynamical effects which might affect the location of this molecule.
  However the prediction of further heavy-quark symmetry partners cannot
  be done solely on the basis of symmetry and requires additional
  information.
  We propose to use the one boson exchange model to fill this gap,
  in which case we will be able to predict or discard
  the existence of other partner states.
  Besides the isoscalar $2^{++}$ $D^*\bar{D}^*$ bound state, we correctly
  reproduce the location and quantum numbers of the isovector
  hidden-bottom $Z_b(10610)$ and $Z_b(10650)$ molecular candidates.
  We also predict the hidden-bottom $1^{++}$ $B^*\bar{B}^*$ and
  $2^{++}$ $B^*\bar{B}^*$ partners of the $X(3872)$, in agreement
  with previous theoretical speculations, plus a series of other states.
  The isoscalar, doubly charmed $1^+$ $D D^*$ and $D^* D^*$ molecules
  and their doubly bottomed counterparts are likely to bind,
  providing a few instances of explicitly exotic systems.
\end{abstract}

\maketitle

\section{Introduction}

Hadronic molecules were conjectured four decades ago from a direct analogy
to the deuteron and the nuclear forces that bind
it~\cite{Voloshin:1976ap,DeRujula:1976qd}.
The idea is that heavy hadrons can exchange light mesons,
such as the pion, the sigma, the rho and the omega,
in the same way as nucleons do.
In a few cases the interaction might be strong enough as to bind
the hadrons into a molecule~\cite{Tornqvist:1991ks,Tornqvist:1993ng,Ericson:1993wy}.
This conjecture has become particularly relevant after the experimental
discovery of the $X(3872)$ by the Belle collaboration~\cite{Choi:2003ue},
which is usually considered to be a $1^{++}$ $D^*\bar{D}$
bound state~\cite{Tornqvist:2003na,Voloshin:2003nt,Braaten:2003he}.
Besides the closeness of the $X(3872)$ to the open charm threshold $D^0 D^{0*}$,
the most convincing evidence of the molecular nature of the $X(3872)$ is
its isospin breaking decays into $J/\Psi \, 2\pi$ and $J/\Psi \, 3\pi$~\cite{Choi:2011fc},
which are easily explained in the molecular picture~\cite{Gamermann:2009fv,Gamermann:2009uq},
but not if the $X(3872)$ is a more compact object~\cite{Hanhart:2011tn}.
However a direct confirmation of the molecular nature of the $X(3872)$
requires precise measurements of
its $D^0 \bar{D}^0\gamma$ and $D^0 \bar{D}^0\pi^0$
decays~\cite{Voloshin:2003nt,Fleming:2007rp,Guo:2014hqa}.
The discovery of the $X(3872)$ was preceded by detection of
the (probably molecular) $D_{s0}^*(2317)/D_{s1}(2460)$
mesons~\cite{Aubert:2003fg,Besson:2003cp},
and has been followed by the observation of other molecular candidates,
in particular the $Z_b$'s~\cite{Belle:2011aa,Garmash:2014dhx},
the $Z_c$'s ($Z_c(3900)$~\cite{Ablikim:2013mio, Liu:2013dau}
and $Z_c(4020)$~\cite{Ablikim:2013wzq, Ablikim:2014dxl}),
and the $P_c(4450)$~\cite{Aaij:2015tga},
which has been recently discovered to consist of two nearby peaks,
the $P_c(4440)$ and $P_c(4457)$, and to have a lighter partner,
the $P_c(4312)$~\cite{Aaij:2019vzc}.

Most of the known molecular candidates are suspected to be bound states of
heavy hadrons, i.e. hadrons containing a heavy quark.
As a consequence their interactions are constrained
by heavy-quark symmetry.
This by itself is able to explain a few interesting properties of
the spectrum of the molecular candidates~\cite{AlFiky:2005jd,Voloshin:2011qa,Mehen:2011yh,Valderrama:2012jv,Nieves:2012tt,Guo:2013sya,Guo:2013xga}.
Heavy-quark symmetry can be divided into heavy-quark spin symmetry (HQSS:
the interaction among heavy hadrons are independent of heavy-quark spin),
heavy flavour symmetry (HFS: the interactions among heavy hadrons are
independent of the heavy-quark flavour) and heavy antiquark-diquark
symmetry (HADS: a heavy antiquark behaves as a heavy diquark pair).
In a few instances heavy-quark symmetry can be used to predict the existence
of unobserved partners of known molecular states.
From a molecular $X(3872)$ 
it is possible to deduce the existence of the $X(4012)$ or
$X_2$~\cite{Valderrama:2012jv,Nieves:2012tt} --- a $2^{++}$
$D^* \bar{D}^*$ partner of the $X(3872)$ --- and a series of
triply heavy pentaquark-like molecules~\cite{Guo:2013xga}.
From heavy-quark symmetry and the assumption
that the $P_c(4450)$~\cite{Aaij:2015tga}
is a $J^P = \tfrac{3}{2}^-$ $\bar{D}^* \Sigma_c$ molecule,
we expect the existence of a
$\tfrac{5}{2}^-$ $\bar{D}^* \Sigma_c^*$
partner state --- which we can call the $P_c(4515)$ in reference
to its expected mass --- plus a few triply heavy hexaquark-like
molecules~\cite{Liu:2018zzu}.
Now with the discovery of new $P_c(4312)$ and the two peak structure
of the $P_c(4450)$~\cite{Aaij:2019vzc} we can effectively
determine the seven possible S-wave heavy meson-baryon
molecules~\cite{Liu:2019tjn,Xiao:2019aya}, including
the previously mentioned $\tfrac{5}{2}^-$ $\bar{D}^* \Sigma_c^*$ state.
Yet, with the exception of the doubly charmed pentaquark family,
heavy-quark symmetry alone is in general not able to determine
the full molecular spectrum by itself and has to be supplemented
with additional information about hadron dynamics.

The one boson exchange (OBE) model~\cite{Machleidt:1987hj,Machleidt:1989tm},
besides having played a central role in the seminal speculations
about the existence of hadron molecules,
is able to provide this missing information about
the hadron-hadron interactions.
In this model the potential between two hadrons is the consequence of
the exchange of a series of light mesons ($\pi$, $\sigma$,
$\rho$ and $\omega$) that provide the necessary dynamics
for binding.
The OBE model is not completely free of ambiguities though:
for making concrete predictions, a form factor and a cutoff are required
to regularize the unphysical short-range behaviour of the light mesons.
If we limit ourselves to qualitative predictions, then it is not necessary
to determine $\Lambda$: it will be enough to have $\Lambda \sim 1\,{\rm GeV}$,
i.e. the natural scale for light hadrons.
But if we want quantitative predictions the specific choice of
the cutoff $\Lambda$ is important.
One of the observations we make in this work is that the cutoff $\Lambda$
can be effectively determined from the condition of reproducing
the binding energy of a known molecular candidate,
e.g. the $X(3872)$.
After determining the cutoff $\Lambda$ with this condition,
we can explore how the OBE model applies to the particular cases of
the heavy meson-meson and heavy meson-antimeson systems and
what predictions are to be expected.
This framework is an adaptation of the {\it renormalized OBE} ideas
of Ref.~\cite{Cordon:2009pj}, which also represents the most
important conceptual innovation of the present work
with respect to previous applications of the OBE model to
heavy hadron molecules, see Refs.~\cite{Liu:2008tn,Yang:2011wz,Sun:2011uh}
as representative examples.
We find that the twin hidden-bottom isovector $Z_b(10610)$ and $Z_b(10650)$
resonance --- $Z_b$ and $Z_b'$ for short --- are correctly reproduced
by the OBE model.
The hidden-bottom partner of the $X(3872)$ is also predicted
in agreement with the previous literature,
from T\"ornqvist~\cite{Tornqvist:1993ng}
onwards.
A series of additional molecular states are predicted, particularly 
in the hidden-bottom sector, which we will discuss later.

The manuscript is structured as follows: in Section \ref{sec:hqss} we will
review the application of heavy-quark spin symmetry to
the heavy meson-antimeson system.
In Section \ref{sec:obe} we will present the details of the one boson exchange
model as applied to the heavy meson-antimeson system.
In Section \ref{sec:pre} we determine the cutoff in the OBE model from the
condition of reproducing the $X(3872)$ as a hadronic molecule, from which
we predict in turn the full spectrum of heavy meson-antimeson molecules.
Finally we present our conclusions in Section \ref{sec:conclusions}.

\section{Heavy Quark Spin Symmetry}
\label{sec:hqss}

We first review the consequences of HQSS for heavy meson molecules.
The quark content of the heavy mesons is $Q \bar{q}$.
If the heavy-light quark-antiquark pair is in S-wave,
the total angular momentum of the heavy meson is $J = 0, 1$.
The $J=0$ and $J=1$ mesons are denoted as $P$ and $P^*$, respectively.
The fields of the $P$ and $P^*$ heavy mesons can be combined 
into a single (heavy-quark spin symmetric) superfield~\cite{Falk:1992cx},
the non-relativistic version of which is
\begin{eqnarray}
  H = \frac{1}{\sqrt{2}}\,
  \left[ P + \vec{P}^* \cdot \vec{\sigma} \right] \, , \label{eq:H-nr}
\end{eqnarray}
where $H$ is a 2x2 matrix and $\vec{\sigma}$ refers to the Pauli matrices.
From the non-relativistic superfield $H$ we can easily construct 
the contact-range Lagrangian for the heavy meson-meson interaction.
If we consider interactions that do not contain derivatives of the heavy
meson fields, the most general Lagrangian will be~\cite{AlFiky:2005jd}
\begin{eqnarray}
  \mathcal{L}_{4H} &=& 
  C_a {\rm Tr}\left[ H^{\dagger} H \right] {\rm Tr}\left[ H'^{\dagger} H' \right]
  \nonumber \\
  &+&
  C_b {\rm Tr}\left[ H^{\dagger} \sigma_i H \right]
  {\rm Tr}\left[ H'^{\dagger} \sigma_i H' \right] \, ,
  \label{eq:HH-EFT-LO}
\end{eqnarray}
where we use $H$ and $H'$ to denote heavy mesons of different flavour.
Notice that we are ignoring isospin or flavour quantum numbers
in the Lagrangian above.
This is actually very powerful, because without HQSS there will be a total of
six independent S-wave interactions which are reduced to two.
If we particularize for a heavy meson-antimeson pair $H = Q \bar{q}$
and $H' = \bar{H} = \bar{Q} q$, we are required to have well-defined C-parity.
In this case, the contact-range non-relativistic potential for S-wave reads
\begin{eqnarray}
  V(0^{++}, P\bar{P}) &=& C_a \, , \\ \nonumber \\
  V(1^{++}, P^*\bar{P}) &=& C_a + C_b \, , \\
  V(1^{+-}, P^*\bar{P}) &=& C_a - C_b \, , \\ \nonumber \\
  V(0^{++}, P^*\bar{P}^*) &=& C_a - 2 C_b \, , \\
  V(1^{+-}, P^*\bar{P}^*) &=& C_a - C_b \, , \\
  V(2^{++}, P^*\bar{P}^*) &=& C_a + C_b \, . 
\end{eqnarray}
This potential contains two interesting patterns.
The first pattern is that the $1^{+-}$ $P^* \bar{P}$ and $P^* \bar{P}^*$
potentials are identical, which implies that these two molecules should
have the same binding energy.
This might explain why the $Z_c$, $Z_c'$ and $Z_b$, $Z_b'$ resonances
come in pairs~\cite{Bondar:2011ev,Mehen:2011yh}
(if they happen to be molecules).
The second pattern is the $1^{++}$ $P^* \bar{P}$ and
$2^{++}$ $P^* \bar{P}^*$ potentials, which are identical.
According to this pattern, if the $1^{++}$ $D^* \bar{D}$ system binds
then the $2^{++}$ $D^* \bar{D}^*$ should also
bind.
In particular, if the $X(3872)$ is a molecule,
there should be a $2^{++}$ molecule with a similar
binding energy~\cite{Valderrama:2012jv,Nieves:2012tt}.
This molecule, the $X(4012)$, has not been observed experimentally yet
and remains theoretical.
{
Besides, other theoretical models predict the $2^{++}$ partner of
the $X(3872)$ to have a different mass that is not necessarily
close to the $D^* \bar{D}^*$ threshold~\cite{Molina:2009ct,Baru:2016iwj}.}

The interesting thing about the contact-range Lagrangian
of Eq.~(\ref{eq:HH-EFT-LO}) is that it is all we need
to describe heavy meson-(anti)meson molecules
with a reasonable degree of accuracy.
The reason for this is that the Lagrangian of Eq.~(\ref{eq:HH-EFT-LO})
can be interpreted as the leading order Lagrangian of an effective
field theory (EFT) for heavy meson interactions~\cite{Valderrama:2012jv}.
Within the EFT framework the heavy meson interaction is divided
into a long- and a short-range part.
The short-range part contains everything with a range of the order of $1/M$
(with $M = 0.5-1.0\,{\rm GeV}$, the typical hadronic energy scale)
or shorter, while the long-range part contains interactions
with a range larger than that figure.
From this definition the long-range piece only contains pion exchanges,
while the short-range piece represents the exchange of all other light
mesons ($\sigma$, $\rho$, $\omega$), the contribution of which
can be effectively encapsulated in a contact-range Lagrangian.
It happens that for heavy meson-(anti)meson systems pion exchanges are
usually {\it subleading}, i.e. their impact in the description of these
systems is not as important as the Lagrangian of Eq.~(\ref{eq:HH-EFT-LO}).

This leaves the contact-range Lagrangian of Eq.~(\ref{eq:HH-EFT-LO})
as the leading order EFT for heavy meson molecules.
However the EFT description has a problem: there are two free parameters,
or four once we consider that the different isospin channels
have different couplings.
The EFT framework does not provide information
about the couplings $C_a$ and $C_b$,
which have to be determined from existing physical information.
For instance, if the $X(3872)$ is indeed an isoscalar $1^{++}$ $D^*\bar{D}$
molecule, from the condition of reproducing the binding energy of
the $X(3872)$ we can determine the linear combination
\begin{eqnarray}
  V_X = C_{0a} + C_{0b} \, ,
\end{eqnarray}
where we have added a subscript to the couplings
to indicate the isospin channel:
$C_{Ia}$ and $C_{Ib}$ with $I=0$ for the $X(3872)$.
Analogously, if the $Z_b$ and $Z_b'$ are isovector $1^{+-}$ $B^*\bar{B}$ and
$B^*\bar{B}^*$ molecules, we can determine another combination of couplings 
\begin{eqnarray}
  V_Z = C_{1a} - C_{1b} \, ,
\end{eqnarray}
in exactly the same way (notice that we did not mention the $Z_c$'s because
their potential is identical to that of the $Z_b$'s
owing to HFS~\cite{Guo:2013xga}).

Here lies the problem we want to address in this manuscript:
besides the $X(3872)$ and the $Z_b$'s there are
no other clear molecular candidates. 
We simply cannot fix the four leading order couplings and determine
the full spectrum of heavy meson-antimeson molecules.
For that reason we have to resort to a phenomenological model if we want
to effectively predict the heavy meson molecular spectrum.
The phenomenological model we will use here is the OBE model.

\section{The One Boson Exchange Model}
\label{sec:obe}

In this section we explain the OBE model
as applied to the heavy meson molecules.
In the OBE model, the interaction between two hadrons is a direct
consequence of the exchange of light mesons.
In its most basic version these light mesons are the pion,
the $\sigma$, the $\rho$ and the $\omega$.
The OBE potential provided the first accurate description of
the nuclear force~\cite{Machleidt:1987hj,Machleidt:1989tm},
and the seminal idea for the first conjectures
about the existence of heavy hadron molecules~\cite{Voloshin:1976ap}.
The OBE potential has its limitations too and
there have been frequent discussions about
the coupling constants of the mesons, in particular regarding
the short-range piece of the OBE potential which is usually
dominated by $\rho$ and $\omega$ exchange.
From SU(3)-flavour symmetry and the OZI rule we expect
$g_{\omega NN} \simeq 3 g_{\rho NN}$, but a good description of
the nuclear scattering data usually requires
$g_{\omega NN} > 3 g_{\rho NN}$.
Nowadays, owing to the conceptual frameworks provided by renormalization and
effective field theory, we understand that these problems are derived
from the fine-tuning of the nuclear forces,
see Ref.~\cite{Cordon:2009pj} for a lucid exposition.
Yet the application of the OBE potential to hadronic molecules is
in part less problematic because of its exploratory character.

\subsection{The Lagrangian}

If we use the non-relativistic superfield $H$ defined in Eq.~(\ref{eq:H-nr}),
we can write the interaction Lagrangian between the heavy and light mesons
as follows:
\begin{eqnarray}
    \mathcal{L}_{H H \pi} &=& -\frac{g}{\sqrt{2} f_{\pi}}\,{\rm Tr}
  \left[ {H}^{\dagger} \vec{\sigma} \cdot \nabla ( \vec{\tau} \cdot \vec{\pi})
    H \right] \, , \label{eq:L-pi} \\
  \mathcal{L}_{H H \sigma} &=& g_{\sigma}\,{\rm Tr}\left[
    {H}^{\dagger} \sigma H \right] \, , \label{eq:L-sigma} \\
  \mathcal{L}_{H H \rho} &=& g_{\rho}\,{\rm Tr}\left[
    {H}^{\dagger} \vec{\tau} \cdot \vec{\rho}^{0} H \right]
  \nonumber \\
  &-& \frac{f_{\rho}}{4 M}\,\epsilon_{ijk}\,{\rm Tr}\left[
    {H}^{\dagger} \sigma_k \vec{\tau} \cdot \left( \partial_i \vec{\rho}_j
    - \partial_j \vec{\rho}_i \right) H \right] \, , \label{eq:L-rho} \\
    \mathcal{L}_{H H \omega} &=& -g_{\omega}\,{\rm Tr}\left[
    {H}^{\dagger} {\omega}^{0} H \right]
  \nonumber \\
  &+& \frac{f_{\omega}}{4 M}\,\epsilon_{ijk}\,{\rm Tr}\left[
    {H}^{\dagger} \sigma_{k} \, \left( \partial_i {\omega}_j
    - \partial_j {\omega}_i \right) H \right] \, . \label{eq:L-omega}
\end{eqnarray}
In the equations above $\pi$, $\sigma$, $\rho_{\mu} = (\rho_0, \rho_i)$
and $\omega_{\mu} = (\omega_0, \omega_i)$ represent the light meson fields,
where there is a Lorentz index in the $\rho$ and $\omega$ fields
as they are vector mesons.
The axial coupling of the pion is denoted by $g$, $g_{\sigma}$ is the coupling
to the $\sigma$ meson, while for the vector mesons we have two different
couplings: $g_V$ and $f_V$, with $V = \rho, \omega$.
The coupling $g_V$ and $f_V$ are the strength of the ``electric-type''
and ``magnetic-type'' interactions of the vector mesons,
respectively.
For the magnetic-type term we include a mass $M$,
which is there to make $f_V$ dimensionless.
We will set $M$ to be the $D$ meson mass: $M = m_D = 1.87\,{\rm GeV}$.

\subsection{The OBE Potential}

The OBE potential is written as the sum of the contributions of
the exchanged mesons ($\pi$, $\sigma$, $\rho$ or $\omega$)
\begin{eqnarray}
  V = \zeta \, V_{\pi} + V_{\sigma} + V_{\rho} + \zeta \, V_{\omega} \, ,
\end{eqnarray}
where $\zeta = \pm 1$ is a sign which we use to distinguish
between the heavy meson-meson and meson-antimeson cases.
We use the convention:
\begin{eqnarray}
  \zeta = +1 && \quad\mbox{for $H\bar{H}$} \, , \\
  \zeta = -1 && \quad\mbox{for $H H$} \, ,
\end{eqnarray}
i.e. the sign is positive for the heavy meson-antimeson case,
which is the most commonly studied case in the context of
heavy hadron molecules.
The contribution of each light meson in momentum space is
\begin{eqnarray}
  V_{\pi}(\vec{q})
  &=& \eta\,\vec{\tau}_1 \cdot \vec{\tau}_2\,\frac{g^2}{2 f_{\pi}^2}\,
  \frac{\vec{a}_1 \cdot \vec{q}\, \vec{a}_2 \cdot \vec{q}}
       {{\vec{q}\,}^2 + \mu_{\pi}^2}
  \, , \label{eq:pi} \\
  V_{\sigma}(\vec{q}) &=& -\frac{g_{\sigma}^2}{{\vec{q}\,}^2 + m_{\sigma}^2} \,
  \label{eq:sigma} \\
  V_{\rho}(\vec{q}) &=& \vec{\tau}_1 \cdot \vec{\tau}_2\, \Big[
    \frac{g_{\rho}^2}{{\vec{q}\,}^2 + m_{\rho}^2} \nonumber \\
    &-&
    \eta\,\frac{f_{\rho}^2}{4 M^2}\,
    \frac{(\vec{a}_1 \times \vec{q}) \cdot
      (\vec{a}_2 \times \vec{q})}{{\vec{q}\,}^2 + \mu_{\rho}^2} \Big]
   \, , \label{eq:rho} \\
   V_{\omega}(\vec{q}) &=& -\frac{g_{\omega}^2}{{\vec{q}\,}^2 + m_{\omega}^2}
   \nonumber \\
    &+&
   \eta\,\frac{f_{\omega}^2}{4 M^2}\,
   \frac{(\vec{a}_1 \times \vec{q}) \cdot
      (\vec{a}_2 \times \vec{q})}{{\vec{q}\,}^2 + \mu_{\omega}^2}
  \, , \label{eq:omega}
\end{eqnarray}
where $\eta = \pm 1$ is a sign, which we will define later, and 
$\vec{a}_1$ and $\vec{a}_2$ vectors that depend on whether
we are considering the $P\bar{P}$, $P^*\bar{P}$/$P\bar{P}^*$ or
$P^*\bar{P}^*$ systems.
The convention for $a_i$ (with $i=1,2$ the vertex) is the following
\begin{eqnarray}
  \vec{a}_i = 0\,\, && \quad \mbox{for a $P \to P$ vertex} , \\
  \vec{a}_i = \vec{\epsilon}_i \,\, &&
  \quad \mbox{for a $P \to P^*$ vertex} , \\
  \vec{a}_i = \vec{\epsilon}_i^* \,\, &&
  \quad \mbox{for a $P^* \to P$ vertex} , \\
  \vec{a}_i = \vec{S}_i && \quad \mbox{for a $P^* \to P^*$ vertex} ,
\end{eqnarray}
with $\vec{\epsilon}_i$ the polarization vector of the $P^*$ meson
and $\vec{S}_i$ the spin-1 matrices.
The convention for the sign $\eta$ is different depending on whether
we are in the heavy meson-meson or heavy meson-antimeson system.
For the heavy meson-meson case we have
\begin{eqnarray}
  \eta = +1 && \quad \mbox{for the $P^* {P} + P^*{P}$ potential} , \\
  \eta = -1 && \quad \mbox{for the $P^* {P} - P^*{P}$ potential} , \\
  \eta = +1 && \quad \mbox{for the $P^* {P}^*$ potential} \, ,
\end{eqnarray}
depending on whether we have a symmetric or antisymmetric $P P^*$ configuration.
For the heavy meson-antimeson case we have
\begin{eqnarray}
  \eta = +1 && \quad \mbox{for the $C=(-1)^L$ $P^* \bar{P}$ potential} , \\
  \eta = -1 && \quad \mbox{for the $C=(-1)^{L+1}$ $P^* \bar{P}$ potential} , \\
  \eta = +1 && \quad \mbox{for the $P^* \bar{P}^*$ potential} \, ,
\end{eqnarray}
where $C$ refers to the C-parity of the heavy meson-antimeson system
and $L$ to the orbital angular momentum.
Notice that for the piece of the potential that is multiplied by $\eta$
we do not use the mass $m$ of the light meson, but the effective mass $\mu$
which is defined as
\begin{eqnarray}
  \mu^2 = m^2 - \Delta_Q^2  && \quad \mbox{for the $PP^*$/$P^*P$ case} , \\
  \mu^2 = m^2 && \quad \mbox{for the $P^* P^*$ case} ,
\end{eqnarray}
where for $Q=c$ we have that $\Delta_{c} = m_{D}^* - m_D$ is the mass splitting
between the $D$ and $D^*$ charmed mesons.
The reason is that these potentials imply a vertex in which the $P$ heavy meson
transitions into a $P^*$ heavy meson and vice versa for the other vertex.
For the vector mesons the difference between $\mu$ and $m$ in the
$PP^*$/$P^*P$ potential is really small, and we will simply take
the approximations $\mu_{\rho} \simeq m_{\rho}$ and
$\mu_{\omega} \simeq m_{\omega}$.
For the pion and the charmed mesons $D$ and $D^*$ we have instead
that $m_{\pi} \simeq \Delta_c$. In this case we will make
the simplification $\mu_{\pi} = 0$ for the $DD^*/D^*D$ potential.

The coordinate space potential is obtained from Fourier-transforming
the potentials of Eqs.~(\ref{eq:pi}-\ref{eq:omega}),
in which case we arrive at
\begin{eqnarray}
  V_{\pi}(\vec{r}) &=&
  -\eta\,\vec{\tau}_1 \cdot \vec{\tau}_2\,\frac{g^2}{6 f_{\pi}^2}\,\Big[
    - \vec{a}_1 \cdot \vec{a}_2\,\delta(\vec{r})
    \nonumber \\ && \quad
    + \vec{a}_1 \cdot \vec{a}_2\,\mu_{\pi}^3\,W_Y(\mu_{\pi} r)
    \nonumber \\ && \quad
    + S_{12}(\vec{r})\,\mu_{\pi}^3\,W_T(\mu_{\pi} r) \Big] \, , \\
  V_{\sigma}(\vec{r}) &=& -{g_{\sigma}^2}\,m_{\sigma}\,W_Y(m_{\sigma} r)
  \, , \\
  V_{\rho}(\vec{r}) &=& \vec{\tau}_1 \cdot \vec{\tau}_2\,\Big[
    {g_{\rho}^2}\,m_{\rho}\,W_Y(m_{\rho} r) \nonumber \\
    && \quad + \eta\frac{f_{\rho}^2}{4 M^2}\,\Big(
    -\frac{2}{3}\,\vec{a}_1 \cdot \vec{a}_2 \, \delta(\vec{r})
    \nonumber \\ && \qquad
    +\frac{2}{3}\,\vec{a}_1 \cdot \vec{a}_2 \, \mu_{\rho}^3 \, W_Y(\mu_{\rho} r)
    \nonumber \\ && \qquad
    -\frac{1}{3}\,S_{12}(\hat{r})\, \mu_{\rho}^3 \, W_T(\mu_{\rho} r) \,\,
    \Big) \, \Big] \, , \\
    V_{\omega}(\vec{r}) &=& 
    -{g_{\omega}^2}\,m_{\omega}\,W_Y(m_{\omega} r) \nonumber \\
    && \quad - \eta\frac{f_{\omega}^2}{4 M^2}\,\,\Big(
    -\frac{2}{3}\,\vec{a}_1 \cdot \vec{a}_2 \, \delta(\vec{r})
    \nonumber \\ && \qquad
    +\frac{2}{3}\,\vec{a}_1 \cdot \vec{a}_2 \,
    \mu_{\omega}^3 \, W_Y(\mu_{\omega} r)
    \nonumber \\ && \qquad
    -\frac{1}{3}\,S_{12}(\hat{r})\, \mu_{\omega}^3 \, W_T(\mu_{\omega} r) \,\,
    \Big) \, , 
\end{eqnarray}
where the functions $W_Y(x)$ and $W_T(x)$ are defined as
\begin{eqnarray}
  W_Y(x) &=& \frac{e^{-x}}{4\pi x} \, , \\
  W_T(x) &=& \left( 1 + \frac{3}{x} + \frac{3}{x^2} \right)
  \,\frac{e^{-x}}{4\pi x} \, .
\end{eqnarray}

\subsection{Form Factors}

The momentum space potentials of Eqs.~(\ref{eq:pi}-\ref{eq:omega}) is
computed under the assumption of point-like particles.
The finite-size of the mesons can be modeled by means of a form factor
\begin{eqnarray}
  V_M(\vec{q}, \Lambda) = V_M(\vec{q})\,F^2(q, m, \Lambda)\,
  \, ,
\end{eqnarray}
where the subscript $M = \pi, \sigma, \rho, \omega$ refers to the light meson
that is being exchanged, $m$ to the mass of the exchanged meson,
and $F$ to the form factor.
Here we will use a multipolar parametrization of
the type~\footnote{{
    This parametrization, which is the most commonly used
  for the OBE model, has been put into question
  in Ref.~\cite{Ozpineci:2013qza} for the vector mesons.
  The reason is that in the hidden gauge formalism the Weinberg-Tomozawa
  terms are perfectly saturated from vector meson exchange, but this only
  happens with a form factor that does not suppress
  the off-shell coupling of the vector mesons to hadrons.
  However, from the point of view of standard chiral perturbation theory,
  vector meson exchange does not saturate the Weinberg-Tomozawa terms
  but their subleading corrections~\cite{Ecker:1988te}, in which case
  the off-shell suppression of a multipolar form factor
  becomes a very welcome feature.}
}
\begin{eqnarray}
  F(q, m, \Lambda)=
  {\left( \frac{\Lambda^{2}-m^2}{\Lambda^{2}-{q}^2} \right)}^n
  \label{Eq:FF} \, ,
\end{eqnarray}
with $q^2 = q_0^2 - \vec{q}\,^2$ the 4-momentum of the exchanged meson
and where $n$ is the power of the multipolar form factor:
for $n=1$ we have a monopolar form factor,
for $n=2$ a dipolar one, etc.
In principle each of the light mesons can have a different form factor
and cutoff, as happens in the OBE model as applied in the two-nucleon system.
But this is only possible if there are plenty of experimental data to fit.
This is not the case for hadronic molecules and thus we will simply choose
to use the same form factor --- a monopolar form factor ($n=1$) --- and
the same cutoff for each of the light mesons.

The inclusion of a form factor can be taken into account
with the following changes in the coordinate space potential
\begin{eqnarray}
  \delta(r) &\to& m^3\,d(x,\lambda) \, , \\
  W_Y(x) &\to& W_Y(x, \lambda) \, , \\
  W_T(x) &\to& W_T(x, \lambda) \, , 
\end{eqnarray}
with $\lambda = \Lambda / m$.
For a monopolar form factor the functions $d$, $W_Y$ and $W_T$ read
\begin{eqnarray}
  d(x, \lambda) &=& \frac{(\lambda^2 - 1)^2}{2 \lambda}\,
  \frac{e^{-\lambda x}}{4 \pi} \, , \\
  W_Y(x, \lambda) &=& W_Y(x) - \lambda W_Y(\lambda x) \nonumber \\ && -
  \frac{(\lambda^2 - 1)}{2 \lambda}\,\frac{e^{-\lambda x}}{4 \pi} \, , \\
  W_T(x, \lambda) &=& W_T(x) - \lambda^3 W_T(\lambda x) \nonumber \\ && -
  \frac{(\lambda^2 - 1)}{2 \lambda}\,\lambda^2\,
  \left(1 + \frac{1}{\lambda x} \right)\,\frac{e^{-\lambda x}}{4 \pi} \, .
\end{eqnarray}
For form factors of higher polarity we refer to Appendix \ref{app:pol}.

\subsection{Couplings}

\begin{table}[!h]
\begin{tabular}{|ccc|}
  \hline\hline
  Light Meson  & $I^{G}\,(J^{PC})$  & M (MeV) \\
  \hline
  $\pi$ & $1^{-}$ $({0}^{-+})$ & 138 \\
  $\sigma$ & $0^{+}$ $({0}^{++})$ & 600 \\
  $\rho$ & $1^{+}$ $({1}^{--})$ & 770 \\
  $\omega$ & $0^{-}$ $({1}^{--})$ & 780 \\
  \hline \hline \hline
  Heavy Meson & $I (J^P)$ & M (MeV) \\
  \hline
  $D$ & $\frac{1}{2}(0^-)$ & 1867 \\
  $D^*$ & $\frac{1}{2}(1^-)$ & 2009 \\
  $B$ & $\frac{1}{2}(0^-)$ & 5279 \\
  $B^*$ & $\frac{1}{2}(1^-)$ & 5325 \\
  \hline \hline
\end{tabular}
\caption{Masses and quantum numbers of the light mesons
  of the OBE model ($\pi$, $\sigma$, $\rho$, $\omega$)
  and the heavy mesons ($D$, $D^*$, $B$, $B^*$)
}
\label{tab:masses}
\end{table}

\begin{table}[!h]
\begin{tabular}{|cc|}
  \hline \hline
  Coupling  & Value for $P$/$P^*$ \\
  \hline
  $g$ & 0.60 \\
  $g_{\sigma}$ & 3.4 \\
  $g_{\rho}$ & 2.6 \\
  $g_{\omega}$ & 2.6 \\
  $\kappa_{\rho}$ & 4.5 \\
  $\kappa_{\omega}$ & 4.5 \\
  $M$ & 1867 \\
  \hline \hline
\end{tabular}
\caption{Couplings of the light mesons of the OBE model
  ($\pi$, $\sigma$, $\rho$, $\omega$) to the heavy meson fields.
  For the magnetic-type coupling of the $\rho$ and $\omega$ vector mesons
  we have used the decomposition
  $f_{V} = \kappa_{V}\,g_{V}$, with $V = \rho, \omega$.
  $M$ refers to the mass scale involved in the magnetic-type couplings.
}
\label{tab:couplings}
\end{table}

The OBE potential depends on the pion axial coupling $g$, the sigma
coupling $g_{\sigma}$, the vector meson electric- and magnetic-type
couplings $g_V$ and $f_V$ with $V=\rho, \omega$ and the mass scale $M$.
For the pion axial coupling we choose
\begin{eqnarray}
  g = 0.6 \, ,
\end{eqnarray}
which is compatible within errors with the experimental determination
$g_1 = 0.59 \pm 0.01 \pm 0.07$ from the $D^* \to D \pi$
decay~\cite{Ahmed:2001xc,Anastassov:2001cw}.
For the sigma coupling, we determine it from the nucleon-nucleon-sigma
coupling in the non-linear sigma model~\cite{GellMann:1960np}
($g_{\sigma NN} = \sqrt{2}\,M_N / f_{\pi} \simeq 10.1$)
and the quark model~\cite{Riska:2000gd} relation:
\begin{eqnarray}
  g_{\sigma} = \frac{1}{3} g_{\sigma NN} \simeq 3.4 \, .
\end{eqnarray}
From Sakurai's universality~\cite{Sakurai:1960ju}
we expect the electric-type $\rho$ coupling to be
\begin{eqnarray}
  g_{\rho} = \frac{m_{\rho}}{2 f_{\pi}} \simeq 2.9 \, ,
\end{eqnarray}
though this is merely a first approximation. For instance,
Casalbuoni et al.~\cite{Casalbuoni:1992dx}
suggested instead
\begin{eqnarray}
  g_{\rho} = \beta \frac{m_{\rho}}{2 f_{\pi}} \simeq 2.6 \, ,
\end{eqnarray}
where $\beta = 0.9$.
We adopt this second estimation for the $\rho$ coupling, which is closer to
the lattice QCD calculation of Ref.~\cite{Detmold:2012ge}:
$g_{\rho} = 2.6 \pm 0.1 \pm 0.4$ in the heavy quark limit.
For the magnetic-type $\rho$ coupling, we also follow Casalbuoni
et al.~\cite{Casalbuoni:1992dx} (which apply vector meson dominance
to the weak decays of the charmed mesons), in which case we obtain
\begin{eqnarray}
  f_{\rho} = 4 \lambda M \frac{m_{\rho}}{2 f_{\pi}} \simeq 11.7
  \quad \mbox{for} \quad M = 1.87 \, {\rm GeV} \, ,
\end{eqnarray}
where $\lambda = 0.6 \pm 0.1 \,{\rm GeV}^{-1}$. The apparently
large value of $f_{\rho}$ is a consequence of taking $M$ equal 
to the $D$ meson mass, instead of a more natural scale.
Finally the couplings to the $\omega$ meson can be deduced
from the ones of the $\rho$ meson, plus SU(3) flavour and
the OZI rule, which lead us to
\begin{eqnarray}
  g_{\omega} = g_{\rho} \quad \mbox{and} \quad f_{\omega} = f_{\rho} \, .
\end{eqnarray}
Alternatively, these two relations can also be derived from writing
the Lagrangian for the interaction between heavy and vector mesons
with SU(3) flavour indices and the vector meson nonet.

\subsection{Wave Functions and Partial Wave Projection}

The general wave function for a two heavy meson system takes the form
\begin{eqnarray}
  | \Psi \rangle = \Psi_{J M}(\vec{r)} | I M_I \rangle\, ,
\end{eqnarray}
where $| I M_I \rangle$ refers to the isospin wave function
and $\Psi_{J M}$ to the spin and spatial wave function.
For the isospin wave function we simply couple the isospin of
the two particles; the only subtlety is the isospin convention
for antiparticles (if we are dealing with a hadron-antihadron system),
which can be consulted in Ref.~\cite{Liu:2018bkx} for the heavy antimeson case.
The $\Psi_{J M}$ piece of the wave function
can be written as a partial wave sum
\begin{eqnarray}
  \Psi_{JM}(\vec{r}) = \sum_{LS} \psi_{L S\,J}(r) | {}^{2S+1}L_J \rangle \, .
\end{eqnarray}
We use the spectroscopic notation ${}^{2S+1}L_J$ for denoting a partial
wave with total spin $S$, orbital angular momentum $L$ and
total angular momentum $J$.
The precise definition is
\begin{eqnarray}
  |{}^{2S+1}L_{J}\rangle &=& \sum_{M_{S},M_L}
  \langle L M_L S M_S | J M \rangle \, | S M_S \rangle \, Y_{L M_{L}}(\hat{r})
  \, , \nonumber \\
\end{eqnarray}
where $\langle L M_L S M_S | J M \rangle$ are the Clebsch-Gordan coefficients,
$| S M_S \rangle$ the spin wavefunction and $Y_{L M_L}(\hat{r})$
the spherical harmonics.
For the $PP$ and $P\bar{P}$ system, the spin wave function is trivial
\begin{eqnarray}
 | S M_S (PP) \rangle = | 0 0 \rangle \, ,
\end{eqnarray}
as we are dealing with spin-0 mesons.
For the $PP^*$/$P^*P$ and $P\bar{P}^*$/$P^*\bar{P}$ system,
only one of the heavy mesons have spin
\begin{eqnarray}
 | S M_S (PP^*) \rangle = | 1 M_S \rangle \, .
\end{eqnarray}
For the $P^*P^*$ and $P^* P^*$ systems we have
\begin{eqnarray}
  | S M_S (P^* P^*) \rangle = \sum_{M_{S1},M_{S2}} &&
  \langle 1 M_{S1} 1 M_{S2} | S M_S \rangle \nonumber \\ &\times& 
  | 1 M_{S1} \rangle \, | 1 M_{S2} \rangle \, , 
\end{eqnarray}
with $| 1 M_{S1} \rangle$, $| 1 M_{S2} \rangle$
are the spin wavefunction of particles $1$ and $2$.

The partial wave projection of the OBE potential depends
on the matrix elements of the $\vec{a}_1 \cdot \vec{a}_2$
and $S_{12}$ operators, which are in turn independent of $J$ and $M$
\begin{eqnarray}
  \langle S' L' J' M' | {\bf O}_{12} | S L J M \rangle &=&
  \delta_{J J'} \delta_{M M'} \, {\bf O}^J_{S L, S' L'} \, ,
\end{eqnarray}
where ${\bf O}_{12} = \vec{a}_1 \cdot \vec{a}_2$ or $S_{12}$.
The specific matrix elements of the spin-spin and tensor operators
can be consulted in Table \ref{tab:tensor} for all the molecular
configurations that contain an S-wave (i.e. the ones that
are more likely to bind).

\begin{table*}[t]
\begin{tabular}{|c|c|c|c|c|}
\hline\hline
Molecule & Partial Waves & $J^P$ & $\vec{a}_1 \cdot \vec{a}_2 $ &
$S_{12} = 3\,\vec{a}_1 \cdot \hat{r} \, \vec{a}_2 \cdot \hat{r} - \vec{a}_1 \cdot \vec{a}_2$
\\ \hline
$D \bar{D}$ & $^1S_0$ & $0^+$  & 0 & 0 \\ \hline
$D^* \bar{D}$ & $^3S_1$-$^3D_1$ & $1^+$ & 
$\left(\begin{matrix}
1 & 0 \\
0 & 1
\end{matrix}\right)$  & $\left(\begin{matrix}
0 & -\sqrt{2} \\
-\sqrt{2} & 1%
\end{matrix}\right)$\\  \hline
$D^* \bar{D}^*$ & $^1S_0$-$^5D_0$  & $0^+$ & 
$\left(\begin{matrix}
-2 & 0 \\
0 & 1 \\
\end{matrix}\right)$& $\left(\begin{matrix}
0 & -\sqrt{2} \\
-\sqrt{2} & -2 \\
\end{matrix}\right)$  \\ \hline
$D^* \bar{D}^*$ & $^3S_1$-$^3D_1$ & $1^+$ & 
$\left(\begin{matrix}
-1 & 0 \\
0 & -1 \\
\end{matrix}\right)$& $\left(\begin{matrix}
0 & \sqrt{2} \\
\sqrt{2} & -1 \\
\end{matrix}\right)$  \\ \hline
$D^* \bar{D}^*$ & $^1D_2$-$^5S_2$-$^5D_2$-$^5G_2$ & $2^+$ & 
$\left(\begin{matrix}
-2 & 0 & 0 & 0 \\
 0 & 1 & 0 & 0 \\
 0 & 0 & 1 & 0 \\
 0 & 0 & 0 & 1
\end{matrix}\right)$& $\left(\begin{matrix}
0 & -\sqrt{\frac{2}{5}} & \frac{2}{\sqrt{7}} & -6\,\sqrt{\frac{3}{35}} \\
-\sqrt{\frac{2}{5}} & 0 & \sqrt{\frac{14}{5}} & 0 \\
\frac{2}{\sqrt{7}} & \sqrt{\frac{14}{5}} &
\frac{3}{7} & \frac{12}{7\sqrt{5}} \\
-6\,\sqrt{\frac{3}{35}} & 0 & \frac{12}{7\sqrt{5}} & -\frac{10}{7} \\ 
\end{matrix}\right)$  \\
\hline\hline
\end{tabular}
\centering \caption{Matrix elements of the spin-spin and
  tensor operator for the partial waves we are considering
  in this work.} \label{tab:tensor}
\end{table*}

\section{Predictions of Molecular States}
\label{sec:pre}

\subsection{The $X(3872)$ as a Renormalization Condition}

The predictions of the OBE potential depend
on the form factor cutoff $\Lambda$.
For a soft cutoff the OBE potential is too weak to form bound states,
while for a hard cutoff the OBE potential is too strong,
leading to overbinding or to spurious bound states.
The physical interpretation of the cutoff $\Lambda$ is that
it represents the finite size of the hadrons.
For the particular case of a multipolar form factor, the cutoff $\Lambda$
is expected to be larger than the masses of the exchanged light mesons
but not considerably larger than the natural hadronic scale,
about $1-2\,{\rm GeV}$ give or take~\footnote{
We notice that form factors of a higher polarity prefer larger values
of $\Lambda$: for a monopolar form factor the ideal cutoff is
in the vicinity of $\Lambda \simeq 1\,{\rm GeV}$, while
for a dipolar $\Lambda \simeq 1.5\,{\rm GeV}$.}.

For the heavy meson-antimeson system it is possible to uniquely
determine $\Lambda$ from the mass of the $X(3872)$.
For concreteness we will consider the $X(3872)$ in the isospin symmetric limit,
where it is an isoscalar $D^* \bar{D}$ bound state with positive
C-parity $C=+1$ and a binding energy of about $4\,{\rm MeV}$.
Notice that the $4\,{\rm MeV}$ figure comes from the difference between
the location of the $X(3872)$ pole and the $D D^*$ threshold
for the isospin averaged masses of the charmed mesons.
With this condition and the parameters of
Tables \ref{tab:masses} and \ref{tab:couplings},
the OBE potential generates the $X(3872)$ pole for the cutoff
\begin{eqnarray}
  \Lambda = \Lambda_X = 1.01\,{\rm GeV} \, ,
\end{eqnarray}
which agrees with our expectations of a natural size cutoff~\footnote{
  In a previous work~\cite{Liu:2018bkx}
  we obtained $\Lambda = 1.04^{+0.18}_{-0.10}\,{\rm GeV}$,
  which is a bit higher.
  The reason for the difference is that in Ref.~\cite{Liu:2018bkx}
  the effective pion mass for the one pion exchange piece of the OBE potential
  was taken to be $\mu_{\pi} = 138\,{\rm MeV}$ instead of $\mu_{\pi} = 0$.
  It is interesting to notice that the difference is indeed small,
  which is consistent with the idea that one pion exchange is
  perturbative, as expected in the charm meson-antimeson system
  from the analysis of Ref.~\cite{Valderrama:2012jv}.
}.
For comparison purposes, we notice that for the deuteron taken as
a neutron-proton bound state with a binding energy $B_2 = 2.2\,{\rm MeV}$
we obtain $\Lambda_d = 0.86\,{\rm GeV}$,
which is of the same order of magnitude.

\subsection{Error Estimations}

With the OBE potential and the cutoff determined
from the renormalization condition,
we are ready to compute the spectrum of heavy meson-(anti)meson  molecules.
The calculation of the spectrum will be affected
by uncertainties that have to be estimated.
The most important source of uncertainty is the OBE potential itself:
with the exceptions of the pion and rho couplings, $g$ and $g_{\rho}$,
the other couplings of the OBE potential are not particularly well known.
For instance $g_{\sigma}$ is derived from the quark model, for which there are no
reliable error estimations, though a $30\%$ looks like a sensible figure;
$g_{\omega}$ has been derived from SU(3)-flavour symmetry and the OZI rule,
where at least a $20\%$ error is to be expected.
These uncertainties will propagate into the calculation of the binding energies.
Considering the error of each of the coupling constants
separately is possible (by means of Monte Carlo techniques,
for instance), but cumbersome.
Instead we will assign an overall relative error $\delta_{\rm OBE}$
for the OBE potential, that is
\begin{eqnarray}
  V = V_{\rm OBE}\,(1 \pm \delta_{\rm OBE}) \, ,
\end{eqnarray}
where we take $\delta_{\rm OBE} = 30\%$.
If we consider the $X$ channel, this will be the only error source
we will consider, i.e.
\begin{eqnarray}
  V_X = V_{X, \rm OBE}\,(1 \pm \delta_{\rm OBE}) \, ,
\end{eqnarray}
which obviously propagates into the determination of the cutoff $\Lambda_X$.
In particular we arrive to
\begin{eqnarray}
  \Lambda_X = 1.01^{+0.18}_{-0.10}\,{\rm GeV} \, .
  \label{eq:Lambda_X_err}
\end{eqnarray}
Now if we consider molecular states different than the $X$, we notice that
the OBE potential for these molecules has been derived from the assumption
that heavy-quark symmetry applies. It happens that heavy-quark symmetry
is not exact, but has an uncertainty.
For that we include an additional, independent error source
\begin{eqnarray}
  V &=& V_{\rm OBE}\,(1 \pm \delta_{\rm OBE})\,(1 \pm \delta_Q) \, ,
\end{eqnarray}
where $\delta_Q$ is the relative size of the expected violation of
heavy-quark symmetry.
In addition the factor $(1 \pm \delta_{\rm OBE})$ is identical to
the one that we have previously taken for $V_X$.
With this, the calculation of the binding energy is trivial:
with $V$ and $\Lambda = \Lambda_X$ we determine the binding energy $B_2$
of the two heavy meson system.
Finally we sum in quadrature the two errors in $B_2$, the one propagated
from $(1 \pm \delta_{\rm OBE})$ and the second one propagated
from $(1 \pm \delta_Q)$.

Besides the two-body binding energy $B_2$, we will also compute the scattering
length $a_2$ of the two-meson systems. In the absence of bound states,
the scattering length is useful to determine the strength of
the interaction, particularly if it is close to binding.
If we define the binding momentum as $\gamma_2 = \sqrt{2 \mu B_2}$, with
$\mu$ the reduced mass of the two-body system and $B_2$ its binding
energy, we have the relation
\begin{eqnarray}
  a_2 = \frac{1}{\gamma_2}\,\left[ 1 + \mathcal{O}
    (\frac{\gamma_2}{m_{\pi}})\right] \, ,
\end{eqnarray}
which is valid for $\gamma_2 \ll m_{\pi}$, where we compare with the pion mass
$m_{\pi}$ because it is the longest range contribution to the OBE potential.
For $B_2 \to 0$ the scattering length diverges: $a_2 \to \infty$.
When the bound state disappears but the attraction in the two-body system is
still sizable, the scattering length will be large (with respect to the
range of the pion) and negative: $a_2 > -1/m_{\pi} \sim -1.4\,{\rm fm}$.
We notice that for the $D^*\bar{D}$ and $D^* D$ systems the effective pion mass
is close to zero, $\mu_{\pi} \simeq 0$. This in turn means that the tensor
force from one pion exchange effectively behaves as a $1/r^3$ potential at
large distance, which are known for not having a well-defined scattering
length. For this reason we will not calculate the $D^*\bar{D}$
and $D^* D$ scattering lengths.
For the calculation of the scattering length in settings with several
coupled partial waves, we refer to Ref.~\cite{PavonValderrama:2005ku}
which deals with the two nucleon system, but where all the formalism
can be easily translated to the two heavy meson system.
The error estimations for the scattering length will be done exactly
as the ones for the two-body binding energy.

Finally we comment that there is another important source of (unknown)
uncertainty in the calculation of $\Lambda_X$: the impact of nearby charmonia.
The $X(3872)$ is thought to be predominantly molecular
as deduced from the isospin breaking decays of the $X$ into
$J/\Psi \pi^{+} \pi^{-}$ and $J/\Psi \pi^{+} \pi^{-} \pi^0$~\cite{Choi:2011fc}.
This branching ratio is naturally explained if the $X(3872)$
is molecular~\cite{Gamermann:2009fv,Gamermann:2009uq},
but not if it is a charmonium state~\cite{Hanhart:2011tn}.
But the $X(3872)$ also decays into a charmonium and a photon,
$\Gamma(X \to J/\Psi \gamma)$ and
$\Gamma(X \to \Psi(2S) \gamma)$~\cite{Aaij:2014ala},
which suggests the existence of $c\bar{c}$ components
at shorter distances~\cite{Swanson:2004pp}.
It is important to notice that only a small short-range $c\bar{c}$ component
is required to explain the radiative charmonium decays~\cite{Dong:2009uf},
i.e. the $X(3872)$ remains to be mostly molecular~\cite{Guo:2014taa}.
It has also been argued that the coupled channel dynamics
between the $D^*\bar{D}$ degrees of freedom and the $\chi_{1}(2P)$
charmonia might provide additional attraction
to the system~\cite{Cincioglu:2016fkm}.
This additional attraction means that the actual form factor cutoff
$\Lambda_X$ that is necessary to bind the $X(3872)$ might be
a bit lower than the value we use here.
However it is difficult to estimate how much lower it is, as this depends
on what is the $c\bar{c}$ probability for the $X(3872)$
{(check also Ref.~\cite{Kang:2016jxw} for a more recent analysis of
the degree of compositeness in the $X(3872)$)}.
For that reason we will not include this effect in our calculations and
simply consider the $X(3872)$ to be a pure molecular state
for simplicity.

\subsection{The Isoscalar Heavy Meson-Antimeson System}

\begin{table}[!ttt]
\begin{tabular}{|cccccc|}
\hline\hline
state  & $I$ & $J^{PC}$ & $a_2$ (fm) & $B_2$ (MeV) & $M$ (MeV) \\
  \hline
  $D \bar D$ & $0^+$ & $0^{++}$ & $-2.1^{+1.7}_{-2.8}$ & - & -\\ \hline
  $D^* \bar D$ & $0^+$ & $1^{++}$ & N/A & $4$ & $3872$ \\
  $D^* \bar D$ & $0^+$ & $1^{+-}$ & N/A & - & - \\ \hline
  $D^* {\bar D}^*$ & $0^+$ & $0^{++}$ & $-1.5^{+0.5}_{-1.0}$ & - & - \\
  $D^* {\bar D}^*$ & $0^+$ & $1^{+-}$ & $-2.0^{+0.9}_{-2.7}$ & - & - \\
  $D^* {\bar D}^*$ & $0^+$ & $2^{++}$ &
  $+2.8^{+4.5}_{-0.8}$ & $4^{+17}_{-6}$ & $4013$ \\
  \hline\hline
state  & $I$ & $J^{PC}$ & $a_2$ (fm) & $B_2$ (MeV) & $M$ (MeV) \\
  \hline
  $B \bar B$ & $0^+$ & $0^{++}$ & $+1.3^{+0.9}_{-0.5}$ & $9^{+13}_{-9}$ & $10550$
  \\ \hline
  $B^* \bar B$ & $0^+$ & $1^{++}$ & $+0.6 \pm 0.3$ & $51^{+45}_{-38}$ & $10553$ \\
  $B^* \bar B$ & $0^+$ & $1^{+-}$ & $+1.5^{+1.2}_{-0.9}$
  & $7^{+12}_{-7}$ & $10595$ \\ \hline
  $B^* {\bar B}^*$ & $0^+$ & $0^{++}$ & $+1.2^{+1.7}_{-1.2}$ & $7^{+21}_{-9}$
  & $10643$ \\
  $B^* {\bar B}^*$ & $0^+$ & $1^{+-}$ & $+1.5^{+1.1}_{-0.9}$ & $8^{+12}_{-7}$
  & $10642$ \\
  $B^* {\bar B}^*$ & $0^+$ & $2^{++}$ & $+0^{+1}_{-25}$ & $59^{+51}_{-43}$
  & $10591$ \\
  \hline\hline 
\end{tabular}
\caption{
  Scattering lengths and binding energies of prospective isoscalar
  heavy meson-antimeson molecules.
  The error is a combination of the expected uncertainty of the OBE model
  and of heavy quark symmetry (HQSS and HFS).
  $M$ refers to the predicted mass (the central value) of
  a heavy meson molecule (if it binds).
  }
\label{tab:binding-hidden-X}
\end{table}

We begin by computing the spectrum for the isoscalar hidden charm and
hidden bottom molecules.
The use of the OBE potential (where the cutoff has been determined from
the location of the $X(3872)$) relies on two types of
heavy-quark symmetry: HQSS and HFS.
The expected relative error of heavy-quark symmetry is
$\delta_{Q} \sim \Lambda_{\rm QCD} / m_Q$,
where $\Lambda_{\rm QCD} \sim 200-300\,{\rm MeV}$.
If we particularize for the charm quark mass, $m_c \sim 1.5\,{\rm GeV}$,
the error will be $\delta_{\rm Q} \sim 15\%$.
If we consider the hidden charm molecules, we will only make use of
HQSS and the relative error $\delta_Q$ for the OBE potential
will simply be $\delta_Q = \delta_{\rm HQSS} \sim 15\%$.
If we consider the hidden bottom molecules, we are actually using
both HQSS and HFS: first, HQSS from applying the OBE potential
for hidden charmed molecules different than the $X$ and second,
HFS from applying the same OBE potential in the hidden charm
and hidden bottom sector.
We account for this by adding the two errors in quadrature.
The error comes from using the hidden charm sector
as the starting point in the calculations,
i.e. we take $\delta_{\rm HQSS} = \delta_{HFS} = \delta_{Q} \sim 15\%$.
Now adding both of these errors in quadrature
we get $\delta_Q' = \sqrt{2}\,\delta_{Q} \sim 20\%$
for the hidden bottom sector.

With the OBE potential and the previous error estimations
we arrive to the set of binding energies listed
in Table \ref{tab:binding-hidden-X}.
In the hidden charm sector we obtain that besides the $X(3872)$
the only other state that might survive is its $2^{++}$ partner,
the $X(4012)$ in reference to its expected mass, which was
predicted in Refs.~\cite{Valderrama:2012jv,Nieves:2012tt}.
The uncertainty is large 
\begin{eqnarray}
  B_2(2^{++}, D^*\bar{D}^*) = 4^{+17}_{-6} \, ,
\end{eqnarray}
where the lower error (which is larger than the central value
as a consequence of summing in quadrature the OBE and
heavy-quark symmetry uncertainties) indicates that
the state might very well disappear.
Thus the conclusion that the $X(4012)$ binds is not strong.
In this sense the present work simply reaffirms the previous conclusions of
Ref.~\cite{Nieves:2012tt}, but which larger uncertainties owing to
the uncertainty of using a phenomenological model
instead of an EFT.

The existence of the $X(4012)$ has indeed been extensively discussed
in the literature from different perspectives.
Despite being a clear prediction of HQSS, the $X(4012)$ has not been
experimentally observed yet.
In principle it could be detected from
$e^+ e^{-} \to \psi(nS) \to \gamma X_2$ (with $\psi$ a $1^{--}$ charmonium)
in the $4.4-4.5\,{\rm GeV}$ region~\cite{Guo:2014ura}.
If it is experimentally discarded in the future, a series of possible reasons
for its disappearance have been already studied, such as
the impact of nearby charmonia~\cite{Cincioglu:2016fkm} or
coupled channel dynamics~\cite{Baru:2016iwj}. 
Here we are however inclined to favor the most simple explanation available:
that the natural uncertainty of HQSS in the charm sector is too large
to guarantee the existence of the $X(4012)$.

If we consider the hidden bottom sector, we arrive to the conclusion that
the six possible isoscalar molecules will bind.
The prediction for the $X_{b1}$, the hidden bottom partner of the $X(3872)$,
is that its binding energy is
\begin{eqnarray}
  B_2(X_{b1}) = 51^{+45}_{-38}\,{\rm MeV} \, .
\end{eqnarray}
This is similar to the original calculation
by T\"ornqvist~\cite{Tornqvist:1993ng},
which used the one pion exchange potential with a monopolar form factor
and a cutoff of $\Lambda = 1.2\,{\rm GeV}$,
leading to $B_2(X_{b1}) \sim 45\,{\rm MeV}$.
The seminal manuscript by T\"ornqvist is more exploratory than our calculations,
as it does consider only the longest range piece of the heavy meson-(anti)meson
interaction (i.e. the one pion exchange potential).
Despite these limitations, T\"ornqvist also predicted the other hidden bottom
molecules that appear in Table \ref{tab:binding-hidden-X}, though in general
the predicted binding energies are considerably larger than our results
(with the exception of the $1^{++}$ and $2^{++}$ hidden bottom molecules,
for which the predictions are similar).
Of course this is due to considering one pion exchange only,
in contrast to the exchange of other light mesons.

The comparison with the more recent calculation of
Guo et al.~\cite{Guo:2013xga} is more interesting:
the authors of Ref.~\cite{Guo:2013xga} use a contact-range effective
field theory at leading order to predict the location of
the $X_{b1}$ and $X_{2b}$ states,
where the $X_{b2}$ refers to the $2^{++}$ $B^* \bar{B}^*$ molecule.
The advantage of this approach is that EFTs calculations are amenable to
systematic error calculations, i.e. they are in principle more reliable
than the phenomenological calculations we are using here.
In this regard it is interesting to check that the calculations of
Ref.~\cite{Guo:2013xga} predict the $X_{b1}$ and $X_{b2}$ to bind
in the $B_2 \sim 25-65\,{\rm MeV}$ range, which is compatible
with the results in Table~\ref{tab:binding-hidden-X}.
For a more complete comparison between the OBE model
and heavy meson EFT we refer to Sect.~\ref{subsec:EFT}.

\subsection{The Isovector Heavy Meson-Antimeson System}

\begin{table}[!h]
\begin{tabular}{|cccccc|}
\hline\hline
state & $I^G$ & $J^{PC}$ & $a_2$ (fm) & $B_2$ (MeV) & $M$ (MeV) \\
  \hline
  $D \bar D$ & $1^-$ & $0^{++}$ & $-0.7^{+0.2}_{-0.4}$ & - & -\\ \hline
  $D^* \bar D$ & $1^-$ & $1^{++}$ & N/A & - & - \\
  $D^* \bar D$ & $1^+$ & $1^{+-}$ & N/A & - & - \\ \hline
  $D^* {\bar D}^*$ & $1^-$ & $0^{++}$ & $-1.9^{+1.1}_{-2.5}$ & - & - \\
  $D^* {\bar D}^*$ & $1^+$ & $1^{+-}$ & $-1.1^{+0.4}_{-0.5}$ & - & - \\
  $D^* {\bar D}^*$ & $1^-$ & $2^{++}$ & $-0.6 \pm 0.4$ & - & - \\
  \hline\hline
  state & $I^G$ & $J^{PC}$ & $a_2$ (fm) & $B_2$ (MeV) & $M$ (MeV) \\
  \hline
  $B \bar B$ & $1^-$ & $0^{++}$ & $+4.6^{+\infty (-24)}_{-3.3}$
  & $0^{+4}_{\dagger}$ & $10559$ \\ \hline
  $B^* \bar B$ & $1^-$ & $1^{++}$ & $-35^{+40}_{-\infty (7)}$ & - & - \\
  $B^* \bar B$ & $1^+$ & $1^{+-}$ & $+1.7^{+1.2}_{-0.4}$
  & $5^{+11}_{-6}$ & $10599$ \\ \hline
  $B^* {\bar B}^*$ & $1^-$ & $0^{++}$ & $+1.2^{+0.4}_{-0.2}$ & $15^{+21}_{-15}$
  & $10635$ \\
  $B^* {\bar B}^*$ & $1^+$ & $1^{+-}$ & $+1.7^{+1.1}_{-0.4}$ & $5^{+11}_{-6}$
  & $10645$ \\
  $B^* {\bar B}^*$ & $1^-$ & $2^{++}$ & $+28_{-24}^{+\infty (-5)}$
  & $0^{+1}_{\dagger}$ & $10650$ \\
  \hline\hline
\end{tabular}
\caption{
  Scattering lengths and binding energies of the prospective isovector
  heavy meson-antimeson molecules. Notice that the $Z_b(10610)$ and
  $Z_b(10650)$ molecular candidates are reproduced
  in the OBE model.
  The table reads as Table \ref{tab:binding-hidden-X}, except for the following
  two details:
  the $\dagger$ symbol in the binding energy $B_2$ indicates here that
  the state disappears either from the OBE or
  the heavy-quark symmetry uncertainty alone.
  The $\pm \infty$ error in the scattering length $a_2$ indicates
  that the scattering length can cross infinity/minus infinity
  as a consequence of the appearance or disappearance of
  a bound state; when this happens we include the expected bound of
  the scattering length in parenthesis.
  }
\label{tab:binding-hidden-Z}
\end{table}

Next we consider the isovector hidden charm and hidden bottom molecules.
The OBE potential leads to the spectrum of Table \ref{tab:binding-hidden-Z},
where we can appreciate that there are no hidden charm isovector molecules
while there are a few hidden bottom ones.
In particular the OBE potential predicts two twin $I^G(J^{PC}) = 1^+(1^{+-})$
$B^* \bar{B}$ and $B^* \bar{B}^*$ molecules with a binding energy of
\begin{eqnarray}
  B_2(1^{+-}, B^{(*)} \bar{B}) = 5^{+11}_{-6}\,{\rm MeV} \, .
\end{eqnarray}
Obviously we are tempted to identify this prediction
with the $Z_b(10610)$ and $Z_b(10650)$ resonances
discovered by Belle~\cite{Belle:2011aa,Garmash:2014dhx}.
Owing to their closeness to the $B^*\bar{B}$ and $B^*\bar{B}^*$ thresholds, 
the $Z_b$'s have been proposed
to be molecular~\cite{Bondar:2011ev,Cleven:2011gp}.
The analysis of Ref.~\cite{Cleven:2011gp} suggests a binding energy of
$B_2 = 4.7^{+2.3}_{-2.2}\,{\rm MeV}$ and $0.11^{+0.14}_{-0.06}\,{\rm MeV}$
for the $Z_b$ and $Z_b'$ respectively.
The more recent analysis of Ref.~\cite{Wang:2018jlv}
suggest $B_2 = 0.9-1.7\,{\rm MeV}$ for the $Z_b$
(i.e. a bit closer to threshold than in Ref.~\cite{Cleven:2011gp}),
while the $Z_b'$ could either be slightly bound ($B_2 \sim 0.7\,{\rm MeV}$)
or be a resonance just above the $B^* \bar{B}^*$ threshold.
These numbers are indeed compatible with our results, which give a bit more
confidence to the hypothesis that the $Z_b$'s are molecular.
Besides the $Z_b$'s, there is another possible $B^*\bar{B}^*$ bound state
for the quantum numbers $I^G(J^{PC}) = 1^-(0^{++})$ with a binding energy of
about $B_2 = 15\,{\rm MeV}$ and there are other two configurations
where the molecules might be close to the unitary limit,
i.e. to having a bound state at threshold.
These two configurations are the isovector $1^-(0^{++})$ $B\bar{B}$ and
$1^-(2^{++})$ $B^*\bar{B}^*$ states. But the errors of these two
predictions are sizable and it is impossible to determine
their fate within the OBE model.
For comparison purposes a recent work~\cite{Baru:2019xnh}, which uses
the EFT formalism as applied to heavy meson-antimeson molecules,
predicts that the six isovector hidden-bottom states will appear
either as resonances above their respective two-meson thresholds
for a pionful EFT or as virtual states for a pionless EFT (i.e.
the overall picture is the same but the details are different).

The isovector hidden charm sector is interesting because
the $Z_c(3900)$ and $Z_c(4020)$ resonances are usually
regarded as probable molecular candidates.
This is despite the fact that they are located a few MeV
above the $DD^*$ and $D^* D^*$ thresholds respectively,
which is not the expected location for a standard S-wave bound state.
Yet it is possible to interpret the $Z_c(3900)$ and $Z_c(4020)$ as
resonances in $DD^*$ and $D^* D^*$ scattering.
This hypothesis makes it natural to expect the $Z_c$'s to be above threshold,
but requires a potential that is repulsive at long distances and
attractive at short distances, which is indeed the case.
In the OBE potential for the isovector $J^{PC} = 1^{+-}$ $D \bar{D}^*$ and
$D^* \bar{D}^*$ systems, the pion provides a repulsive long-range
contribution and the sigma an attractive medium- and
short-range contribution.
However the $\rho$ and $\omega$ contributions cancel out perfectly
in the limit where SU(3)-flavour symmetry and the OZI rule are exact.
The $\rho$ and $\omega$ cancellation is problematic because
it leads to a potential that is not strong enough to bind.
In turn this has prompted a few authors to consider the role of
  two-pion exchange in the $Z_c$'s~\cite{Aceti:2014kja,Aceti:2014uea}
  and $Z_b$'s~\cite{Dias:2014pva}, a contribution
  which in the OBE model can be identified
  with the exchange of the $\sigma$ meson.
Caution is advised however: as argued in Ref.~\cite{Albaladejo:2015lob}
the experimental information currently available might not be enough
as to determine whether the $Z_c$'s are bound states, virtual states
or resonances. In Ref.~\cite{Gong:2016hlt} it is argued that both
the bound and virtual state interpretations of the  $Z_c(3900)$
are possible (with a slight preference towards the virtual state).
For this reason what we will check is whether the interaction in the
isovector $J^{PC} = 1^{+-}$ $D \bar{D}^*$ and $D^* \bar{D}^*$ systems
is strong, for instance by looking at the scattering length predictions.
In Table \ref{tab:binding-hidden-Z} it can be appreciated that the scattering
length for the $Z_c'$ channels is $a_2 = -1.1^{+0.4}_{-0.5}\,{\rm fm}$,
while for the $Z_c$ the scattering length is not well-defined
in the $\mu_{\pi} \to 0$ limit we are taking.
This scattering length might be compatible with the existence of a virtual
state (which basically requires a large negative scattering length).
Unfortunately the magnitude of the scattering length is natural (of the order
of the pion range), which means that this virtual state is probably
not observable.
According to this result, it is difficult to accommodate
the $Z_c$'s as pure molecular states.

\begin{table}[!h]
\begin{tabular}{|ccccccc|}
\hline\hline
state & Scenario & $I^G$ & $J^{PC}$ & $a_2$ (fm) & $B_2$ (MeV) & $M$ (MeV) \\
  \hline
  $D \bar D$ & B & $1^-$ & $0^{++}$ & $-0.9^{+0.3}_{-0.2}$ & - & -\\ \hline
  $D^* \bar D$ & B & $1^-$ & $1^{++}$ & N/A & - & - \\
  $D^* \bar D$ & B & $1^+$ & $1^{+-}$ & N/A & - & - \\ \hline
  $D^* {\bar D}^*$ & B & $1^-$ & $0^{++}$ & $-4.4^{+2.7}_{\infty (+36)}$ & - & - \\
  $D^* {\bar D}^*$ & B & $1^+$ & $1^{+-}$ & $-1.7^{+0.7}_{-1.4} $ & - & - \\
  $D^* {\bar D}^*$ & B & $1^-$ & $2^{++}$ & $-0.8 \pm 0.3$ & - & - \\
  \hline\hline
state & Scenario & $I^G$ & $J^{PC}$ & $a_2$ (fm) & $B_2$ (MeV) & $M$ (MeV) \\
  \hline
  $D \bar D$ & C & $1^-$ & $0^{++}$ & $-1.0^{+0.3}_{-0.5}$ & - & -\\ \hline
  $D^* \bar D$ & C & $1^-$ & $1^{++}$ & N/A & - & - \\
  $D^* \bar D$ & C & $1^+$ & $1^{+-}$ & N/A & - & - \\ \hline
  $D^* {\bar D}^*$ & C & $1^-$ & $0^{++}$ & $-12^{+12}_{-\infty (+6)}$ & - & - \\
  $D^* {\bar D}^*$ & C & $1^+$ & $1^{+-}$ & $-2.3^{+1.2}_{-3.2}$ & - & - \\
  $D^* {\bar D}^*$ & C & $1^-$ & $2^{++}$ & $-0.9 \pm 0.3$ & - & - \\
  \hline\hline
  state & Scenario & $I^G$ & $J^{PC}$ & $a_2$ (fm) & $B_2$ (MeV) & $M$ (MeV) \\
  \hline
  $B \bar B$ & B & $1^-$ & $0^{++}$ &
  $+2.5^{+4.2}_{-0.9}$ & $2^{+6}_{-3}$ & $10557$ \\ \hline
  $B^* \bar B$ & B & $1^-$ & $1^{++}$ &
  $+8_{-5}^{+\infty (-11)}$ & $0^{+2}_{\dagger}$ & $10604$ \\
  $B^* \bar B$ & B & $1^+$ & $1^{+-}$
  & $+1.3^{+0.5}_{-0.4}$ & $10^{+17}_{-10}$ & $10594$ \\ \hline
  $B^* {\bar B}^*$ & B & $1^-$ & $0^{++}$
  & $+1.0^{+0.4}_{-0.3}$ & $26^{+32}_{-24}$ & $10624$ \\
  $B^* {\bar B}^*$ & B & $1^+$ & $1^{+-}$ & $+1.3^{+0.5}_{-0.4}$
  & $11^{+16}_{-10}$ & $10639$ \\
  $B^* {\bar B}^*$ & B & $1^-$ & $2^{++}$ & $+6^{+\infty (-15)}_{-4}$
  & $0^{+2}_{\dagger}$ & $10650$ \\
  \hline\hline
  state & Scenario & $I^G$ & $J^{PC}$ & $a_2$ (fm) & $B_2$ (MeV) & $M$ (MeV) \\
  \hline
  $B \bar B$ & C & $1^-$ & $0^{++}$ & $+2.1^{+4.2}_{-0.9}$
  & $3^{+7}_{-4}$ & $10556$ \\ \hline
  $B^* \bar B$ & C & $1^-$ & $1^{++}$ & $+5^{+\infty (-110)}_{-3}$ &
  $1^{+2}_{\dagger}$ & $10603$ \\
  $B^* \bar B$ & C & $1^+$ & $1^{+-}$ & $+1.1^{+0.4}_{-0.3}$ & $14^{+20}_{-14}$
  & $10590$ \\ \hline
  $B^* {\bar B}^*$ & C & $1^-$ & $0^{++}$ & $+0.9^{+0.2}_{-0.4}$ &
  $34^{+39}_{-22}$ & $10616$ \\
  $B^* {\bar B}^*$ & C & $1^+$ & $1^{+-}$ & $+1.1^{+0.4}_{-0.3}$
  & $14^{+20}_{-14}$ & $10636$ \\
  $B^* {\bar B}^*$ & C & $1^-$ & $2^{++}$ & $+5^{+86}_{-3}$ &
  $1^{+3.0}_{\dagger}$ & $10649$ \\
  \hline\hline
\end{tabular}
\caption{
  Scattering lengths and binding energies of prospective isoscalar
  heavy meson-antimeson molecules if the $\rho$ and $\omega$ couplings
  break SU(3)-flavour symmetry.
  The table reads as Tables \ref{tab:binding-hidden-X}
  and \ref{tab:binding-hidden-Z}.
  }
\label{tab:binding-hidden-Z-breaking}
\end{table}

This prompts us to consider the role of SU(3)-flavour symmetry breaking
in the formation of the $Z_c's$ resonances.
If SU(3)-flavour symmetry is broken in the right direction, in particular
by having $g_{\omega} > g_{\rho}$, this will generate additional short-range
attraction that might lead to a larger scattering length or to binding.
We will consider the following three scenarios
\begin{itemize}
\item Scenario A: SU(3)-flavour symmetry and OZI rule are exactly respected,
  i.e. $g_{\omega} = g_{\rho}$ and $f_{\omega} = f_{\rho}$.
\item Scenario B: $g_{\omega} = g_{\rho}$ and $f_{\omega} = f_{\rho}$ are
  moderately broken in the right way as to easen
  the binding of the $Z_c's$.
\item Scenario C: same as scenario B, but now the $g_{\omega} = g_{\rho}$
  and $f_{\omega} = f_{\rho}$ relations are strongly broken. 
\end{itemize}
Scenario A simply corresponds to Table \ref{tab:binding-hidden-Z}, i.e.
we follow the choice of coupling constants that we already made
when discussing the OBE model.
In scenario B we acknowledge that the $g_{\omega} = g_{\rho}$ and
$f_{\omega} = f_{\rho}$ relations can be off by a $\delta_{3} = 20\%$,
while in scenario C the relations will be violated at
the $\delta_{3} = 35\%$ level.
The type of breakdown that renders more probable the existence of the $Z_c$'s
as molecular states is $g_{\omega} > g_{\rho}$ and $f_{\omega} > f_{\rho}$.
In particular we take
\begin{eqnarray}
  g_\omega' = (1+\delta_{3})\,g_{\rm SU(3)} \quad &,& \quad g_\rho' = (1-\delta_{3})\,g_{\rm SU(3)} \, , \\
  f_{\omega}' = (1+\delta_{3})\,f_{\rm SU(3)} \quad &,& \quad f_\rho' = (1-\delta_{3})\,f_{\rm SU(3)} \, ,
\end{eqnarray}
where $g_{\rm SU(3)}$ and $f_{\rm SU(3)}$ refer to the previous values we were using
for $g_{\rho / \omega}$ and $f_{\rho / \omega}$.
With these values we have to calculate $\Lambda_X$ again,
in which case we obtain
\begin{eqnarray}
  \Lambda_X' = 1.03^{+0.20}_{-0.10}\,{\rm GeV} \, ,
\end{eqnarray}
which is curiously identical for the $\delta_{3} = 20\%$ and $35\%$
scenarios and very close to the original $\Lambda_X$ in
the SU(3)-symmetric limit, see Eq.~(\ref{eq:Lambda_X_err})
(this happens because the attraction lost
from the $\rho$ in the $X(3872)$ channel is canceled out by the attraction
gained from the $\omega$).
From $\Lambda_X'$ we can recalculate the full spectrum of isoscalar
hidden charm and hidden bottom molecules, in which case
we arrive at the results of Table \ref{tab:binding-hidden-Z-breaking}.
We notice that for scenario $B$ and $C$ the scattering length in the $Z_c'$
channel increases to $a_2^{(B)} = -1.7^{+0.7}_{-1.4} \,{\rm fm}$ and
$a_2^{(C)} = -2.3^{+1.2}_{-3.2}\,{\rm fm}$, respectively.
The scattering lengths cannot be excluded to be large
once we consider the uncertainty, yet neither
the $Z_c$ or $Z_c'$ bind for sensible values of SU(3) breaking.

\subsection{The Heavy Meson-Meson System}

\begin{table}[!h]
\begin{tabular}{|cccccc|}
\hline\hline
state  & $I^G$ & $J^{P}$ & $a_2$ (fm) & $B_2$ (MeV) & $M$ (MeV) \\
  \hline
  $D D$ & $1^+$ & $0^{+}$ & $-0.4^{+0.1}_{-0.2}$ & - & -\\ \hline
  $D^* D + D D^*$ & $1^+$ & $1^{+}$ & N/A & - & -\\
  $D^* D - D D^*$ & $0^+$ & $1^{+}$ & N/A & $3^{+15}_{-4}$ &
      $3873$ \\ 
  \hline
  $D^* D^*$ & $1^+$ & $0^{+}$ & $-0.4 \pm 0.2$ & - & -\\ 
  $D^* D^*$ & $0^+$ & $1^{+}$ & $4^{+100}_{-2}$ & $2^{+13}_{-3}$ &
      $4015$ \\ 
  $D^* D^*$ & $1^+$ & $2^{+}$ & $-0.6^{+0.4}_{-0.3}$ & - & -\\ 
  \hline \hline
  state  & $I^G$ & $J^{P}$ & $a_2$ (fm) & $B_2$ (MeV) & $M$ (MeV) \\
  \hline
  $B B$ & $1^+$ & $0^{+}$ & $-4.5^{+4.0}_{-\infty (+150)}$ & - & -\\ \hline
  $B^* B + B B^*$ & $1^+$ & $1^{+}$ & $+2.4^{+7.6}_{-0.9}$ & $2^{+8}_{-3}$
  & $10602$ \\
  $B^* B - B B^*$ & $0^+$ & $1^{+}$ & $+0.5^{+0.4}_{-0.8}$
  & $58^{+55}_{-44}$ & $10546$ \\ \hline
  $B^* B^*$ & $1^+$ & $0^{+}$ & $-1.7^{+0.8}_{-1.3}$ & - & -\\ 
  $B^* B^*$ & $0^+$ & $1^{+}$ & $+0.5^{+0.4}_{-0.8}$ & $58^{+56}_{-43}$ & $10592$ \\ 
  $B^* B^*$ & $1^+$ & $2^{+}$ & $+2.4^{+6.3}_{-0.9}$ & $2^{+9}_{-3}$
  & $10648$ \\  \hline
  \hline\hline 
\end{tabular}
\caption{
  Scattering lengths and binding energies of prospective
  heavy meson-meson molecules.
  Owing to the requirement of symmetric wave functions, the spin and isospin
  of the molecules are constrained by the relation $(-1)^{I+S+L+1} = 1$.
  The exception is the $D^*D$/$DD^*$ system, which can appear
  in both isospin configurations,
  though the potential is different for each one.
  The table reads as Tables \ref{tab:binding-hidden-X}
  and \ref{tab:binding-hidden-Z}.
  }
\label{tab:binding-doubly}
\end{table}

For the heavy meson-meson system our results are listed
in Table \ref{tab:binding-doubly}.
The most notable prediction in the doubly charmed sector is
the twin isoscalar $J^P = 1^+$ $D D^*$ and $D^*D^*$ bound states,
which are predicted to have a binding energy of
\begin{eqnarray}
  B_2(1^+, D D^*) &\simeq& 3^{+15}_{-4}\,{\rm MeV} \, , \\
  B_2(1^+, D^* D^*) &\simeq& 2^{+13}_{-3}\,{\rm MeV} \, ,
\end{eqnarray}
where the binding energies are almost identical
as a consequence of HQSS.
There have been speculations about the existence of a doubly charmed
tetraquark-like state with the quantum numbers of
the twin doubly charmed molecules we predict.
In the quark model the location of the ground state of the isoscalar
$J^P = 1^+$ $c {c} \bar{q} \bar{q}$ tetraquark configuration
can vary considerably, being sometimes predicted below~\cite{Carlson:1987hh,Gelman:2002wf,Vijande:2009kj,Junnarkar:2018twb} and
sometimes above~\cite{Karliner:2017qjm,Eichten:2017ffp,Mehen:2017nrh}
the $D D^*$ threshold.
A recent work considers the two pion exchange potential
in the heavy meson-meson system~\cite{Xu:2017tsr},
predicting the isoscalar $J^P = 1^+$ $D D^*$ to be bound
by about $20\,{\rm MeV}$.
In the lattice this tetraquark state has been recently predicted to be
$23 \pm 11\,{\rm MeV}$ below the $D D^*$ threshold~\cite{Junnarkar:2018twb}.
The OBE model prediction indeed reinforces the previous speculations,
though it gives a prediction much closer to threshold.

Owing to HFS, we also predict twin isoscalar $1^+$ $B B^*$ and
$B^* B^*$ molecules with a binding energy of $58\,{\rm MeV}$,
which are the heavy flavour partners of the $D D^*$ and $D^* D^*$
isoscalar molecules we discussed in the previous paragraph.
This is in comparison with Ref.~\cite{Wang:2018atz}, which predicts
the isoscalar $J^P = 1^+$ $B B^*$ and $B^* B^*$ bound states to
have a binding energy of $13\,{\rm MeV}$ and $24\,{\rm MeV}$
respectively, which moderately violate HQSS for S-wave interactions
according to which both states should have similar binding energies.
This suggest that in Ref.~\cite{Wang:2018atz}, which uses
the two-pion exchange potential, binding is maybe due to
the SD-wave transitions induced by the tensor force.
Recently, two lattice QCD calculations~\cite{Junnarkar:2018twb,Leskovec:2019ioa}have predicted the isoscalar
$J^P = 1^+$ $ud \bar b \bar b$ tetraquark to be located at
$143 \pm 34\,{\rm MeV}$~\cite{Junnarkar:2018twb} and
$128 \pm 24 \pm 10 {\rm MeV}$~\cite{Leskovec:2019ioa}
below the $B B^*$ threshold, respectively.
Besides this exotic doubly bottomed tetraquark-like molecules,
there are two other shallow (isovector) molecules,
a $J^P = 1^+$ $B B^*$ and a $2^+$ $B^* B^*$ bound state,
see Table \ref{tab:binding-doubly} for details.

\subsection{Systems with Two Different Flavours}

Finally we consider the $D B$ (charmed-antibottom) and
$D \bar{B}$ (charmed-bottom) family of
heavy meson molecules.
The most interesting is the charm-antibottom sector, which we summarize
in Table \ref{tab:binding-two-flavour-cbbar}.
In particular we predict a series of molecular candidates
close to the unitary limit, i.e. which a scattering length
considerably larger than the range of molecular interaction
(the range of the pion): $m_{\pi}\,a_2 \gg 1$.
The possibility that the $D B$, $D B^*$ and $D^* B$ molecules
might have unnaturally large scattering lengths / form shallow bound state
has been already theorized in Ref.~\cite{Valderrama:2018sap}
from a simple argument involving the heavy-quark spin decomposition
of the heavy meson-antimeson interaction.
The present calculation confirms
the suspicions of Ref.~\cite{Valderrama:2018sap} independently and adds
a few more charm-antibottom molecules which are also close to
the unitary limit: the isoscalar/isovector
$J^P = 0^+$/$1^+$ $D^* B^*$ molecules.
The isoscalar charm-bottom sector is also interesting: it contains a possible
$J^P = 1^+$ $D^* \bar{B}^*$ bound state, which is the HFS partner of the
doubly charmed and doubly bottomed molecule predicted
in Table \ref{tab:binding-two-flavour-cb}.
In addition though the $D\bar{B}$, $D\bar{B}^*$ and $D^*\bar{B}$ and
$2^+$ $D^*\bar{B}^*$ systems do not bind, their scattering lengths
are also remarkably large.

The reason why the appearance of large scattering lengths is particularly
interesting is because of the possibility of finding a few hadronic
systems where {\it universality} happens~\cite{Braaten:2004rn}.
{\it Universality} is the idea that all two-body systems
with large scattering lengths (in comparison with
the characteristic range of their interaction)
can be described in the same way.
A really interesting aspect of {\it universality} manifests
when we consider the type of three-body systems that are
derived from universal two-body systems.
Three-body systems in which the two-body subsystems
are close to the unitary limit can in principle display
the Efimov effect~\cite{Efimov:1970zz},
i.e. the existence of a geometric tower of three-body bound states
where the ratio of the binding energies of a bound state and
the next excited state approaches a constant value.
The Efimov effect has been extensively studied in molecular physics
(for a recent review see~\cite{Naidon:2016dpf}),
where it was experimentally confirmed for the first time 
with cesium atoms~\cite{Kraemer:2006}.
Efimov physics is also known to play an important role
in nuclear physics~\cite{Hammer:2010kp},
for instance in the description of the triton ~\cite{Bedaque:1998kg,Bedaque:1998km,Bedaque:1999ve},
halo nuclei~\cite{Federov:1994cf,Hammer:2017tjm}
and maybe even in the Hoyle state~\cite{Hammer:2008ra}.
Our results strongly indicate that the bottom-bottom-anticharm three meson
system probably is one of the best candidates to find an Efimov trimer
in hadronic physics, as originally suggested
in Ref.~\cite{Valderrama:2018sap}

\begin{table}[!h]
\begin{tabular}{|cccccc|}
\hline\hline
state  & $I$ & $J^{P}$ & $a_2$ (fm) & $B_2$ (MeV) & $M$ (MeV) \\
  \hline
  $D B$ & $0$ & $0^{+}$ & $+14^{+\infty (-4)}_{-15}$ & $0^{+3}_{\dagger}$
  & $7147$ \\ \hline
  $D B^*$ & $0$ & $1^{+}$ & $+14^{+\infty (-4)}_{-14}$ & $0^{+3}_{\dagger}$
  & $7192$ \\
  $D^* B$ & $0$ & $1^{+}$ & $+7^{+\infty (-6)}_{-7}$
  & $0^{+3}_{\dagger}$
  & $7288$ \\ \hline
  $D^* B^*$ & $0$ & $0^{+}$ & $-7_{-\infty (+13)}^{+5}$ & - & -\\ 
  $D^* B^*$ & $0$ & $1^{+}$ & $+270_{-380}^{+\infty (-3)}$ & - & -\\ 
  $D^* B^*$ & $0$ & $2^{+}$ & $+1.4^{+0.5}_{-0.3}$ & $20^{+33}_{-22}$ & $7314$ \\ 
  \hline \hline
  state  & $I$ & $J^{P}$ & $a_2$ (fm) & $B_2$ (MeV) & $M$ (MeV) \\
  \hline
  $D B$ & $1$ & $0^{+}$ & $-1.5^{+0.6}_{-1.1}$ & - & -\\ \hline
  $D B^*$ & $1$ & $1^{+}$ & $-1.5^{+0.6}_{-1.2}$ & - & -\\
  $D^* B$ & $1$ & $1^{+}$ & $-1.7^{+0.7}_{-1.7}$ & - & -\\ \hline
  $D^* B^*$ & $1$ & $0^{+}$ & $+6^{+\infty (-5)}_{-4}$
  & $0^{+8}_{\dagger}$ & $7334$ \\ 
  $D^* B^*$ & $1$ & $1^{+}$ & $-6^{+4}_{-\infty (+10)}$ & - & -\\ 
  $D^* B^*$ & $1$ & $2^{+}$ & $-1.3^{+0.5}_{-0.8}$ & - & -\\
  \hline\hline 
  \end{tabular}
\caption{
  Scattering lengths and binding energies of prospective
  heavy meson-meson molecules in the flavour-exotic charm-antibottom sector.
  The table reads as Tables \ref{tab:binding-hidden-X}
  and \ref{tab:binding-hidden-Z}.
  }
\label{tab:binding-two-flavour-cbbar}
\end{table}

\begin{table}[!h]
\begin{tabular}{|cccccc|}
  \hline \hline
  state  & $I$ & $J^{P}$ & $a_2$ (fm) & $B_2$ (MeV) & $M$ (MeV) \\
  \hline
  $D \bar{B}$ & $0^+$ & $0^{+}$ & $-5^{+5}_{-\infty (+22)}$ & - & -\\ \hline
  $D \bar{B}^*$ & $0^+$ & $1^{+}$ & $-5^{+5}_{-\infty (+21)}$ & - & -\\
  $D^* \bar{B}$ & $0^+$ & $1^{+}$ & $-8^{+8}_{-\infty (+9)}$ & - & -\\ \hline
  $D^* \bar{B}^*$ & $0^+$ & $0^{+}$ & $+1.0^{+0.1}_{-0.3}$ & $69^{+88}_{-65}$
  & $7265$ \\ 
  $D^* \bar{B}^*$ & $0^+$ & $1^{+}$ & $+1.4^{+0.7}_{-0.4}$ & $16^{+33}_{-19}$
  & $7318$ \\ 
  $D^* \bar{B}^*$ & $0^+$ & $2^{+}$ & $-7^{+5}_{-\infty (+21)}$ & - & -\\ 
  \hline\hline
    state  & $I$ & $J^{P}$ & $a_2$ (fm) & $B_2$ (MeV) & $M$ (MeV) \\
  \hline
  $D \bar{B}$ & $1^+$ & $0^{+}$ & $-0.7 \pm 0.4$ & - & -\\ \hline
  $D \bar{B}^*$ & $1^+$ & $1^{+}$ & $-0.7 \pm 0.4$ & - & -\\
  $D^* \bar{B}$ & $1^+$ & $1^{+}$ & $-0.8^{+0.5}_{-0.4}$ & - & -\\ \hline
  $D^* \bar{B}^*$ & $1^+$ & $0^{+}$ & $-0.6^{+0.1}_{-0.2}$ & - & -\\ 
  $D^* \bar{B}^*$ & $1^+$ & $1^{+}$ & $-0.6^{+0.2}_{-0.4}$ & - & -\\ 
  $D^* \bar{B}^*$ & $1^+$ & $2^{+}$ & $-1.7^{+1.1}_{-3.4}$ & - & -\\ 
  \hline\hline 
\end{tabular}
\caption{
  Scattering lengths and binding energies of prospective
  heavy meson-meson molecules in the flavour-exotic charm-bottom sector.
  The table reads as Tables \ref{tab:binding-hidden-X}
  and \ref{tab:binding-hidden-Z}.
  }
\label{tab:binding-two-flavour-cb}
\end{table}

\subsection{Comparison with Heavy Meson EFT}
\label{subsec:EFT}

The OBE potential is a model, by which we mean that it is not clear
how to estimate the reliability of the predictions.
In contrast EFTs are systematically improvable and allow
for reliable error estimations.
Inspired models, like the OBE model, are phenomenologically successful.
This success is not a matter of method, but the outcome of
inspired choices of what to include in the model.
Thus it is not trivial to determine the theoretical error of
the binding energies and scattering lengths that
we have derived from the OBE model, except with a direct
comparison to experimental data.
As a matter of fact the comparison to experiment is there,
with the $Z_b$ and $Z_b'$ resonance being correctly postdicted by the OBE model,
but we are nonetheless limited to these two examples.

There are additional ways to indirectly assess
the reliability of the OBE model.
One possibility is to compare the predictions of the OBE model
with the ones derived from an EFT.
For this we will compare with the EFT for heavy meson molecules
developed in Ref.~\cite{Valderrama:2012jv}, which has been used
in a series of works about heavy meson
molecules~\cite{Nieves:2012tt,Guo:2013xga}.
The EFT of Ref.~\cite{Valderrama:2012jv}, which we will call heavy meson EFT,
is a refinement of previous ideas, in particular the contact theory with HQSS
of Ref.~\cite{AlFiky:2005jd} and X-EFT~\cite{Fleming:2007rp}.
The problem with heavy meson EFT (or with any other EFT) is that
systematicity comes at the price of predictive power:
the EFT formulation of Ref.~\cite{Valderrama:2012jv} contains
four independent couplings for the contact-range potential.
These couplings are free parameters within the EFT and
have to be determined from experimental information,
e.g. from the location of a known hadronic molecule.
It happens that the number of promising heavy meson-antimeson candidates
is limited to the $X(3872)$, the $Z_c$'s and the $Z_b$'s.
If this were not enough, the $Z_c$'s and $Z_b$'s are connected by means of
HQSS and HFS: heavy meson EFT predicts that their contact-range potentials
are identical.
For this reason, of the four parameters of the heavy meson EFT
at leading order --- namely $C_{0a}$, $C_{0b}$, $C_{1a}$ and $C_{1b}$ ---
only two combinations can be determined, which are
\begin{eqnarray}
  V_X = C_{0a} + C_{0b} \quad \mbox{and} \quad V_Z = C_{1a} - C_{1b} \, ,
\end{eqnarray}
from which a limited number of additional predictions can be made.
By comparing these few predictions with the corresponding ones
in the OBE model we can form a better idea about
the reliability of the OBE model.

Another possibility for testing the reliability of the predictions is to compare
the OBE model with itself, by which we mean to compare the predictions
obtained with different form factors but the same
{\it renormalization condition}.
If we choose a dipolar form factor (instead of a monopolar one),
the cutoff for which the $X(3872)$ pole is reproduced changes to
\begin{eqnarray}
  \Lambda_X^D = 1.41^{+0.28}_{-0.15}\,{\rm GeV} \, ,
\end{eqnarray}
which is roughly $\sqrt{2}$ larger than the monopolar
cutoff~\footnote{For the deuteron
with a dipolar form factor we have $\Lambda_d^D = 1.23\,{\rm GeV}$,
also a factor of $\sqrt{2}$ larger than with a monopolar cutoff.}.
If the change of the binding energy predictions with the dipolar form factor
lie within the errors we have estimated (which they do),
this will also point towards the reliability of the model.

\begin{table}[!ttt]
\begin{tabular}{|ccccccc|}
\hline\hline
state  & $I^G$ & $J^{P}$ & $B_{\rm OBE}^M$ & $B^D_{\rm OBE}$
& $B_{\rm EFT}$($\Lambda = 0.5$)
& $B_{\rm EFT}$($\Lambda=1.0$) \\
  \hline
  $D^*\bar{D}$ & $0^+$ & $1^{++}$ & Input & Input & Input & Input \\
  $D^*\bar{D}^*$ & $0^+$ & $2^{++}$ &
  $4^{+17}_{-6}$ & $4^{+14}_{-5}$ & $5^{+5}_{-4}$ & $5^{+12}_{-5}$ \\
  \hline
  $D^* B^*$ & $0^+$ & $2^{++}$ &  $20^{+33}_{-22}$ & $17^{+25}_{-17}$ &
  $12^{+7}_{-6}$ & $26^{+20}_{-16}$ \\
  \hline
  $B^*\bar{B}$ & $0^+$ & $1^{++}$ &
  $51^{+45}_{-38}$ & $41^{+36}_{-29}$ & $24^{+8}_{-9}$ & $65^{+27}_{-25}$ \\
  $B^*\bar{B}^*$ & $0^+$ & $2^{++}$ &
  $59^{+51}_{-43}$ & $46^{+38}_{-32}$ & $24^{+8}_{-9}$ & $66^{+27}_{-25}$ \\
  \hline \hline
  state  & $I^G$ & $J^{P}$ & $B^M_{\rm OBE}$ & $B^D_{\rm OBE}$ & $B_{\rm EFT}$($\Lambda = 0.5$)
  & $B_{\rm EFT}$($\Lambda=1.0$) \\ \hline
  $B^*\bar{B}$ & $1^+$ & $1^{+-}$ & $5^{+11}_{-6}$ & $4^{+9}_{-5}$ & Input & Input \\
  $B^*\bar{B}^*$ & $1^+$ & $1^{+-}$ & $5^{+11}_{-6}$ 
  & $5^{+9}_{-5}$ & $2.1 \pm 2.1$ & $2.1^{+2.5}_{-2.1}$
  \\
  \hline
  state  & $I^G$ & $J^{P}$ & $a^M_{\rm OBE}$ & $a^D_{\rm OBE}$
  & $a_{\rm EFT}$($\Lambda = 0.5$)
  & $a_{\rm EFT}$($\Lambda=1.0$) \\ \hline
  $D^* B^*$ & $1^+$ & $1^{+-}$ & $-6^{+4}_{-\infty (+10)}$ & $-6^{+5}_{-\infty (+10)}$
  & $-8^{+7}_{-\infty (+16)}$ & $-1.2 \pm 0.6$
  \\ \hline
  $D^* \bar{D}$ & $1^+$ & $1^{+-}$ & N/A & N/A & $-1.4^{+0.8}_{-1.0}$ & $-0.4 \pm 0.1$\\ 
  $D^* \bar{D}^*$ & $1^+$ & $1^{+-}$ & $-1.1^{+0.4}_{-0.5}$ & $-1.2^{+0.4}_{-0.5}$
  & $-1.6^{+1.1}_{-1.1}$ & $-0.5^{+0.2}_{-0.1}$ \\
  \hline \hline
\end{tabular}
\caption{
  Comparison of the predictions of the OBE model with different form factors
  (monopolar and dipolar) and with heavy meson EFT.
  $B_{\rm OBE}^M$ and $B_{\rm OBE}^D$ are the binding energy (in ${\rm MeV}$)
  computed from the OBE model and the {\it renormalization condition}
  with a monopolar and dipolar form factor, respectively.
  $B_{\rm EFT}(\Lambda)$ is the binding energy (in ${\rm MeV}$)
  in heavy meson EFT for a given cutoff $\Lambda$ (in ${\rm GeV}$),
  as taken from Ref.~\cite{Guo:2013xga}.
  For the cases in which the system does not bind --- namely the $Z_c$
  and $Z_c'$ channels and their charm-antibottom counterpart ---
  we compute the scattering length instead.
  }
\label{tab:comparison-EFT}
\end{table}

The comparison with the EFT and the dipolar form factor predictions
is shown in Table \ref{tab:comparison-EFT}.
The heavy meson EFT predictions can be divided into two groups:
the predictions derived from the existence of the $X(3872)$
and the ones derived from the $Z_b(10610)$.
We denote which state has been used to determine the EFT couplings
with the term ``Input'' in Table \ref{tab:comparison-EFT}.
The comparison is actually very interesting: the predictions of the OBE model
with a monopolar and dipolar form factor are compatible between themselves
and with the ones from EFT within errors.
Besides, the uncertainties of the OBE model predictions are
in general larger than the EFT ones.
This might indicate two things: (i) that we have overestimated the OBE errors,
though only by a small margin, or
(ii) that the EFT errors have been underestimated, as recently
hypothesized in Ref.~\cite{Baru:2018qkb} based on the impossibility
of formulating a cutoff-independent EFT compatible with HFS
for heavy hadron molecules.
Be it as it may, the similarity of the OBE and EFT predictions suggests
an acceptable degree of reliability.

\section{Discussion and Conclusions}
\label{sec:conclusions}

We have considered the heavy meson-antimeson and heavy meson-meson systems
from the following two assumptions: (i) the OBE model describes their
interaction and (ii) heavy-quark symmetry further constrains
the dynamics of these systems.
The physics of the OBE potential are intuitive and well-motivated,
but there is the limitation that it requires a form factor
and a cutoff for predictions to be possible.
While the choice of form factor is not that important, the choice of a cutoff
is crucial: without a way to reliably determine the cutoff
it is not possible to make concrete predictions.
For determining this cutoff we have used the assumption that the $X(3872)$
is indeed a $D^*\bar{D}$ bound state with quantum numbers $I=0$,
$J^{PC} = 1^{++}$ and a binding energy of about $4\,{\rm MeV}$
in the isospin symmetric limit (where we use the isospin
symmetric limit for simplicity).
From the cutoff determined with this {\it renormalization condition},
predictions in the OBE model are possible.
We also include error estimations for these predictions.

If we consider the isoscalar hidden charm sector,
we find that besides the existence
of the $X(3872)$ it is plausible to expect that the $X_{c2}$ ---
the $I=0$, $J^{PC} = 2^{++}$ $D^* \bar{D}^*$ system --- also binds.
The expected binding energy of the $X_{c2}$ is $B_2 = 4^{+17}_{-6}\,{\rm MeV}$,
where the uncertainty is however too large to guarantee
the existence of this HQSS partner of the $X(3872)$.
The other isoscalar hidden charm molecules are not expected to bind,
even after taking into account the uncertainty of
the OBE model and HQSS.
It is interesting to compare these results with previous explorations.
The possible existence of the $X_{c2}$ was already discussed
in the seminal work of T\"ornqvist~\cite{Tornqvist:1993ng},
which predicted a isoscalar $J^{PC} = 1^{++}$ $D\bar{D}^*$ bound state
(presumably the $X(3872)$) and pointed out that the $J^{PC} = 2^{++}$
$D^*\bar{D}^*$ system was close to binding, requiring only
a bit of extra attraction to bind.
In Ref.~\cite{Nieves:2012tt} the full six possible HQSS partners of
the $X(3872)$ were predicted, though this work indicated that
the predictions depend on a series of assumptions,
with some predictions more reliable than others.
In particular Ref.~\cite{Nieves:2012tt} indicates that the most robust
prediction is that of the $X_2$, which solely relies on the hypothesis
that the $X(3872)$ is molecular.
This is also what we find in our exploration.
It is also worth mentioning the $I=0$, $J^{PC} = 0^{++}$ $D\bar{D}$ system,
which according to theoretical
explorations~\cite{Gamermann:2006nm,Nieves:2012tt}
could also form a shallow molecule, the $X(3700)$.
Here we find a considerable amount of attraction in the isoscalar $D\bar{D}$
system, which has a negative scattering length $a = -2.1^{+1.7}_{-2.8}\,{\rm fm}$,
but no binding within the uncertainties of the OBE model.
But we did not consider coupled channel effects, which mix the $0^{++}$
$D\bar{D}$ and $D^*\bar{D}^*$ systems and lead to additional attraction.
From EFT arguments coupled channel effects are expected to be a small
correction for two heavy meson molecules~\cite{Valderrama:2012jv},
but the $D\bar{D}$ system is close to binding and small effects
could make a difference.
We will not consider coupled channels in this work, but we mention that
a breaking of HQSS by a $40\%$, which is not particularly probable
statistically but not particularly improbable either,
will lead to binding.

In the isoscalar hidden bottom sector the conclusion is that all
the six possible molecules can bind, with the $J^{PC} = 1^{++}$ and $2^{++}$
molecules --- the $X_{b1}$ and $X_{b2}$ --- being the most bound ones
with $B_2 \sim 50-60\,{\rm MeV}$.
We notice that the first prediction of the $X_{b1}$ and $X_{b2}$
--- the hidden bottom partner of the $X(3872)$ --- already
appears in T\"ornqvist~\cite{Tornqvist:1993ng}.
After this a series of theoretical works~\cite{Wong:2003xk,AlFiky:2005jd,Nieves:2011zz,Guo:2013sya}
--- including ours --- have only reinforced this conclusion further.
The only problem is that the $X_{b1}$ has not been detected in experiments.
In this regard Karliner and Rosner~\cite{Karliner:2014lta} have suggested
that the $\chi_{b1}(3P)$ (with a mass $M = 10512\,{\rm MeV}$) might
not be a $J^{PC} = 1^{++}$ bottomonium after all,
but the bottom partner of the $X(3872)$.
The quantum numbers of the $\chi_{b1}(3P)$ indeed coincide with
the $X_{b1}$ and the required binding energy lies within
the error estimations of the OBE model
($M = 10508-10591\,{\rm MeV}$).

The isovector hidden charm sector is also interesting
owing to its connection with the $Z_c(3900)$ and $Z_c(4012)$
molecular candidates.
The application of the OBE model with SU(3)-symmetric couplings leads to
the conclusion that this two molecules do not bind but are probably
virtual states instead, as deduced from the moderately large
negative scattering length.
This is the same conclusion as in Ref.~\cite{Guo:2013sya}.
We point out that even if the $Z_c(3900)$ and $Z_c(4012)$ are assumed
to be molecular, it cannot be determined if they are genuine bound states,
resonances or virtual states from the experimental data,
see Ref.~\cite{Albaladejo:2015lob} for details.
In this regard the SU(3)-symmetric OBE model will be compatible
with the virtual state hypothesis.
If we allow for natural violations of SU(3)-flavour symmetry in the couplings,
the situation is qualitatively the same as before: there is no binding
within the expected theoretical uncertainties.
However we predict larger scattering lengths than in the SU(3)-symmetric
limit, pointing further towards the virtual state hypothesis.

In the isovector hidden bottom sector the $Z_b$'s are correctly
postdicted as bound states, both of them with a binding energy
of $B_2 = 5^{+11}_{-6} \, {\rm MeV}$.
This figure is not far away from other estimations of their binding
energies, for instance the estimations based on the analysis of
the experimental data done in Refs.~\cite{Cleven:2011gp,Wang:2018jlv}.
Besides, finding the $Z_b$'s in the OBE model further substantiates
the idea that they have a sizable molecular component,
as proposed in Refs.~\cite{Bondar:2011ev,Cleven:2011gp}.

Regarding the doubly charmed sector, we find that the $I=0$, $J^P = 1^+$
$D^* D$ and $D^*D^*$ systems form molecules
with binding energies of $B_2 = 3^{+15}_{-4}\,{\rm MeV}$ and
$2^{+13}_{-3}\,{\rm MeV}$, respectively.
This type of hadron with $cc {\bar q} {\bar q}$ quark content has indeed
been predicted in the quark model~\cite{Carlson:1987hh,Gelman:2002wf,Vijande:2009kj,Junnarkar:2018twb,Karliner:2017qjm,Eichten:2017ffp,Mehen:2017nrh}
(as a compact tetraquark, with large uncertainties regarding
its location though), in a molecular model that includes
two-pion exchange~\cite{Xu:2017tsr}
(with a binding energy of $B_2 = 20\,{\rm MeV}$) and
recently in the lattice~\cite{Junnarkar:2018twb},
with $B_2 = 22 \pm 11\,{\rm MeV}$.
The HFS partners in the charm-bottom ($cb {\bar q} {\bar q}$) and
doubly bottom ($bb {\bar q} {\bar q}$) sectors are also predicted,
with binding energies of $B_2 = 15^{+30}_{-20}\,{\rm MeV}$ and
$60^{+60}_{-50}\,{\rm MeV}$, respectively.
The first qualitative prediction of the $QQ {\bar q}{\bar q}$-family of
tetraquark-like molecules (where $Q = b,c$, i.e. a heavy quark)
was made long ago by Manohar and Wise~\cite{Manohar:1992nd}.
The recent lattice calculations of Refs.~\cite{Junnarkar:2018twb,Leskovec:2019ioa} suggest
a binding energy for the isoscalar $J^P = 1^+$ bottom-bottom tetraquark
of $143 \pm 34\,{\rm MeV}$ and $128 \pm 24 \pm 10\,{\rm MeV}$ respectively
(relative to the $B B^*$ thresholds),
while the recent quark-model calculation of Ref.~\cite{Caramees:2018oue}
locates the isoscalar $J^P = 1^+$ charm-bottom tetraquark at
$B_2 = 164\,{\rm MeV}$ with respect to the $D^* \bar{B}^*$ threshold.
Notice that here we are predicting a molecular state instead of
a compact tetraquark.

Of particular interest is the charm-antibottom sector
($c{\bar b} q {\bar q}$-type molecules), where a series of two-body states
with large scattering lengths are predicted.
This in turns points out to the possibility of Efimov physics in
the $BBD$, $BBD^*$, $B B^* D$, $B^*B^* D$ and $B^* B^* D^*$ three body systems,
as previously conjectured in Ref.~\cite{Valderrama:2018sap}.
Besides having a two-body subsystem close to the unitary limit, the family of
bottom-bottom-charm three body systems displays a moderate mass imbalance
between the charm and bottom mesons, which is a factor
that is known to enhance Efimov physics~\cite{Helfrich:2010yr}.
This family of three hadron systems probably provides one of the most
promising systems in which to observe Efimov trimers, which have been
so far only been observed in atomic systems.
For this reason the exploration of the charm-antibottom sector, either
experimentally or in the lattice, is a really interesting subject.

Finally we have tried to determine the reliability of the OBE model
as applied to heavy meson molecules.
Models, in contrast to theories, are not amenable to error estimations
that are fully systematic.
For this reason it is of particular importance to carefully confront
the OBE model predictions with other approaches.
From the experimental point of view, the correct postdiction of the $Z_b$'s
indicates that the OBE model correctly describes the bulk of
the physics of heavy meson molecules.
From the theoretical point of view, we have compared a set of predictions
derived from heavy meson EFT with the ones of the OBE model.
The agreement seems to indicate that the OBE model is reliable.

\section*{Acknowledgments}

We would like to thank Muhammad Naeem Anwar for comments.
This work is partly supported by the National Natural Science Foundation
of China under Grant No. 11735003, the Fundamental
Research Funds for the Central Universities,
the Youth Innovation Promotion Association CAS (No. 2016367)
and the Thousand Talents Plan for Young Professionals.

\appendix
\section{Multipolar Form Factors}
\label{app:pol}

The OBE model generates a singular potential, where the tensor components
of the potential diverge as $1/r^3$ at short distances.
This type of divergence is unphysical
and can be regularized by means of a form factor.
The most common type of form factor for the OBE model is the multipolar
form factor we have written in Eq.~(\ref{Eq:FF}), where depending
on the exponent we talk about a monopolar ($n=1$),
dipolar ($n=2$), etc. form factor.
In principle the exponent $n$ can depend on the meson $M$, though here
we will assume that all the mesons have the same type of form factor.
In general the contribution of meson $M$ to the OBE potential will be
obtained by Fourier-transforming from momentum to coordinate space as
\begin{eqnarray}
  V_M(\vec{r}) = \int \frac{d^3 \vec{q}}{(2 \pi)^3} V_M(\vec{q})\,
  {\left( \frac{\Lambda^{2}-m^2}{\Lambda^{2}-{q}^2} \right)}^{2 n} \, ,
\end{eqnarray}
but as we have seen, this transformation is relatively direct once we
take into account that the contribution of the form factor can be
encapsulated by the substitutions
\begin{eqnarray}
  \delta(r) &\to& m^3\,d(x,\lambda; 2 n) \, , \\
  W_Y(x) &\to& W_Y(x, \lambda; 2 n) \, , \\
  W_T(x) &\to& W_T(x, \lambda; 2 n) \, , 
\end{eqnarray}
with $\lambda = \Lambda / m$ and where we have labeled them with $2 n$,
i.e. with twice the polarity of the form factor for the exchanged meson.
The function $d$ can be evaluated analytically for $k = 2n = 1,2,3, \dots$
(i.e. integer $k$), with
\begin{eqnarray}
  d(x, \lambda; 1) &=& (\lambda^2 - 1)\,
  \frac{e^{-\lambda x}}{4 \pi x} \, , \\
  d(x, \lambda; k \geq 2) &=& 
  \frac{i \, (\lambda^2 - 1)^{k}}{(k-1)!\,2^{k-1}\,\lambda^{2k-3}}\,
  (i \lambda x)^{k-1} \frac{h^{(+)}_{k-2}(i \lambda x)}{4 \pi} \, ,
  \nonumber \\
\end{eqnarray}
where $h_n^{(\pm)}(z) = j_n(z) \pm i y_n(z)$ are the Haenkel spherical functions,
which we have defined in terms of the Bessel spherical functions
$j_n(z)$ and $y_n(z)$.
For the function $W_Y$ we can evaluate it recursively as
\begin{eqnarray}
  W_Y(x, \lambda; 1) &=& W_Y(x) - \lambda W_Y(\lambda x) \, , \\
  W_Y(x, \lambda; k \geq 2) &=& W_Y(x, \lambda; k-1) -
  \frac{d(x,\lambda; k)}{\lambda^2 - 1}  \, ,
\end{eqnarray}
while for $W_T$ we have
\begin{eqnarray}
  W_T(x, \lambda; 1) &=& W_T(x) - \lambda^3 W_T(\lambda x) \, , \\
  W_T(x, \lambda; 2) &=& W_T(x, \lambda; 1) \nonumber \\ &-&
  \frac{(\lambda^2 - 1)}{2 \lambda}\,\lambda^2\,
  \left(1 + \frac{1}{\lambda x} \right)\,\frac{e^{-\lambda x}}{4 \pi} \, , \\
  W_T(x, \lambda; 3) &=& W_T(x, \lambda; 2) \nonumber \\ &-&
  \frac{{(\lambda^2 - 1)}^2}{8 \lambda}\,(\lambda x)\,
  \frac{e^{-\lambda x}}{4 \pi} \, , \\
  W_T(x, \lambda; k \geq 4) &=& W_T(x, \lambda; k-1) \nonumber \\ &-&
  \frac{i\,(\lambda^2 - 1)^{k-1}}{(k-1)!\,2^{k-1}\,\lambda^{2k-7}}\,x^2
  (i \lambda x)^{k-3} \frac{h^{(+)}_{k-4}(i \lambda x)}{4 \pi} \, . \nonumber \\
\end{eqnarray}
A monopolar form factor (on both vertices) corresponds to the $k=2$ solution,
while a dipolar one to the $k=4$ solution.


%

\end{document}